\documentclass[conference]{IEEEtran}
\makeatletter
\def\ps@headings{%
\def\@oddhead{\mbox{}\scriptsize\rightmark \hfil \thepage}%
\def\@evenhead{\scriptsize\thepage \hfil \leftmark\mbox{}}%
\def\@oddfoot{}%
\def\@evenfoot{}}
\makeatother
\usepackage{ifpdf}
\usepackage{hyperref}
\usepackage{cite}
 \usepackage[pdftex]{graphicx}
 \usepackage{epstopdf}
\usepackage{algorithmicx,algpseudocode}
\usepackage{tabularx}
\usepackage{tabularx,ragged2e,booktabs}
\usepackage{algpseudocode}
\usepackage{graphicx}
\newcommand\CONDITION[2]%
\

\algdef{SE}[WHILE]{While}{EndWhile}[1]%
{\algorithmicwhile\ \CONDITION{#1}{\ \algorithmicdo}}%
{\algorithmicend\ \algorithmicwhile}
\algdef{SE}[FOR]{For}{EndFor}[1]%
{\algorithmicfor\ \CONDITION{#1}{\ \algorithmicdo}}%
{\algorithmicend\ \algorithmicfor}
\algdef{S}[FOR]{ForAll}[1]%
{\algorithmicforall\ \CONDITION{#1}{\ \algorithmicdo}}
\algdef{SE}[REPEAT]{Repeat}{Until}{\algorithmicrepeat}[1]%
{\algorithmicuntil\ \CONDITION{#1}{}}
\algdef{SE}[IF]{If}{EndIf}[1]%
{\algorithmicif\ \CONDITION{#1}{\ \algorithmicthen}}%
{\algorithmicend\ \algorithmicif}%
\algdef{C}[IF]{IF}{ElsIf}[1]%
{\algorithmicelse\ \algorithmicif\ \CONDITION{#1}{\ \algorithmicthen}}
 \usepackage{algorithm}
 \algnewcommand\algorithmicinput{\textbf{Input:}}
 \algnewcommand\Input{\item[\algorithmicinput]}
  \algnewcommand\algorithmicoutput{\textbf{Output:}}
  \algnewcommand\Output{\item[\algorithmicoutput]}
\algrenewcommand\algorithmicindent{1.3em}%
 \usepackage{makecell}
 \usepackage{bm,array}
  \usepackage{array}
  \newcolumntype{P}[1]{>{\centering\arraybackslash}p{#1}}
 \usepackage{booktabs}
 
 \usepackage{multirow}
 \usepackage[normalem]{ulem}
\usepackage{amssymb}
\usepackage{mwe}

 \usepackage{ltablex,booktabs}
 \usepackage{tabularx}
 \usepackage{algorithmicx,algpseudocode}
\usepackage{tabularx,ragged2e,booktabs}
\usepackage{algpseudocode}
 \usepackage[utf8]{inputenc}
 \usepackage{enumerate}
\usepackage{caption}
\newcolumntype{b}{X}
\newcolumntype{s}{>{\hsize=.1\hsize}X}
\usepackage{indentfirst}
\usepackage{makebox}
\usepackage{tabu}
\usepackage{booktabs}

\usepackage{amsthm}

\usepackage{subcaption}
  \renewcommand{\tablename}{Table}
 \renewcommand{\figurename}{Fig.}
 \renewcommand{\thetable}{\arabic{table}}
 
\usepackage{subfloat}
\usepackage{stfloats}
\usepackage{float}
\usepackage{authblk}
\usepackage[T1]{fontenc}
\usepackage[utf8]{inputenc}
\usepackage{xcolor}
\definecolor{lightgray}{rgb}{0.93,0.93,0.93}

\usepackage{rotating}
\usepackage{caption}
\usepackage{siunitx}
\usepackage{xcolor}
\usepackage[tikz]{bclogo}
\usepackage[framemethod=tikz]{mdframed}
\usepackage{lipsum}
\usepackage[many]{tcolorbox}

\newtheoremstyle{mystyle}
{}
{}
{\itshape}
{}
{\bfseries}
{.}
{ }
{}

\theoremstyle{mystyle}

 \usepackage{multicol}
 \let\OLDthebibliography\thebibliography
 \renewcommand\thebibliography[1]{
 	\OLDthebibliography{#1}
 	\setlength{\parskip}{0pt}
 	\setlength{\itemsep}{0pt plus 0.3ex}
 }
 { \begin{itemize}
 		\setlength{\itemsep}{0pt}
 		\setlength{\parskip}{0pt}
 		\setlength{\parsep}{0pt}     }
 	{ \end{itemize}                  } 

\usepackage[english]{babel}
\usepackage{fancyhdr}
\usepackage{lastpage}
\usepackage[utf8]{inputenc }
\usepackage{alertmessage}

\usepackage{mdwmath}

\usepackage[font=footnotesize]{caption}
\usepackage{geometry} 
\usepackage{newunicodechar}

\usepackage{fixltx2e}
\usepackage{stfloats}
\usepackage{url}

\usepackage{graphicx}
\usepackage{float}

\newtcolorbox{hintBox}{textmarker,
    borderline west={6pt}{0pt}{yellow},
    colback=yellow!10!white}
\newtcolorbox{importantBox}{textmarker,
    borderline west={6pt}{0pt}{red},
    colback=red!10!white}
\newtcolorbox{noteBox}{textmarker,
    borderline west={6pt}{0pt}{green},
    colback=green!10!white}

\usepackage[multiple]{footmisc}

    \title{Responsible Design Patterns for Machine Learning Pipelines}

\author[1,2,3]{Saud Hakem Al Harbi \thanks{A.saud.al-harbi@polymtl.ca}}
\author[2]{Lionel Nganyewou Tidjon }
\author[1]{Foutse Khomh, Senior Member, IEEE }

\affil[1]{Department of Computer Engineering, Polytechnique Montreal, Quebec, Canada}
\affil[2]{Department of Engineering, CertKOR AI, Montreal, Quebec, Canada}
\affil[3]{Tabiah University, Medina, Saudi Arabia }

\begin{document}
\maketitle

\begin{abstract}

Integrating ethical practices into the AI development process for artificial intelligence (AI) is essential to ensure safe, fair, and responsible operation. AI ethics involves applying ethical principles to the entire life cycle of AI systems. This is essential to mitigate potential risks and harms associated with AI, such as algorithm biases. To achieve this goal, responsible design patterns (RDPs) are critical for Machine Learning (ML) pipelines to guarantee ethical and fair outcomes. In this paper, we propose a comprehensive framework incorporating RDPs into ML pipelines to mitigate risks and ensure the ethical development, deployment, and post-deployment of AI systems. Our framework comprises new responsible AI design patterns for ML pipelines and a bias mitigation pattern validated through a survey of AI ethics and data experts on real-world scenarios. The framework guides data scientists and policymakers to implement ethical practices in AI development, deploy and monitor responsible AI systems in production.

\end{abstract}
\vspace{2mm} 


\section{Introduction}

Artificial intelligence (AI) development is gaining prominence, requiring the integration of ethical practices into developing AI systems, particularly in the context of machine learning pipelines. As a result, responsible AI and ethical practices in AI development have emerged as crucial considerations in recent years. As AI technologies play an increasingly central role in our lives, adopting AI ethics becomes imperative. AI ethics entails the application of ethical principles throughout the lifecycle of AI systems, ensuring their operation in a safe, fair, and responsible manner. Extensive research has been conducted on this subject, with numerous studies focused on identifying optimal practices and approaches for incorporating ethical considerations into AI development \cite{Alshehhi2022, Daza2022, DeSanctis2020, Lu2022d, nabavi2022five, Ryan2021}.

The growing recognition of AI's potential risks and harms is a critical driver for adopting AI ethics. These risks encompass algorithmic biases that can lead to unfair treatment of individuals or groups, privacy breaches arising from collecting and processing personal data, and the potential for AI systems to cause harm to people or the environment. Organizations and governments worldwide are progressively embracing AI ethics principles and frameworks to address these risks\footnote{\href{www.ibm.com/topics/ai-ethics\#toc-ethical-ai-dPA2LUn3}{IBM}}\footnote{\href{https://ai.google/}{AI Google}}\footnote{\href{www.industry.gov.au/publications/australias-artificial-intelligence-ethics-framework/australias-ai-ethics-principles}{www.industry.gov.au}}.

The European Commission recently proposed new AI regulations that aim to ensure that AI systems are transparent and accountable and operate in a safe and ethical manner \cite{EU2021}. In addition, the IEEE Global Initiative on Ethics of Autonomous and Intelligent Systems has developed a set of ethical principles for AI that are widely recognized and used \cite{IEEE2018}. Incorporating AI ethics into developing and deploying AI systems can have numerous benefits, such as enhancing safety, protecting human rights, and fostering trust and confidence in AI technologies. However, challenges such as more standardization and potential conflicts between different ethical principles exist. Despite these challenges, AI ethics adoption is essential for ensuring that AI technologies are developed and used responsibly and ethically. By considering ethical principles during the design, development, and deployment of AI systems, organizations and governments can ensure that these systems are safe, fair, and beneficial \cite{Lu2022b, Vakkuri2021, Khan2022, Daza2022}. The Saudi Data and Artificial Intelligence Authority (SDAIA) has introduced seven AI ethics principles for public feedback. These principles provide a helpful framework for integrating ethical considerations into all stages of AI system development.\cite{saudi-ai-ethics}. Compliance is a crucial aspect of the AI ethics pipeline, as it ensures that the development and deployment of AI systems adhere to relevant laws, regulations, and ethical standards. Compliance in the AI ethics pipeline can be achieved by incorporating ethical principles into the design and development process, conducting regular audits and assessments to identify and mitigate potential ethical risks, and implementing robust governance and accountability frameworks to oversee the AI system's operations.
Additionally, compliance can ensure that the AI system's outputs and decisions are transparent, explainable, and fair to all stakeholders involved. One example of compliance in the AI ethics pipeline is the General Data Protection Regulation (GDPR), a regulation in the European Union (EU) that governs the processing and handling of personal data. Compliance with GDPR requires AI developers to ensure that their AI systems handle personal data following GDPR's principles, such as ensuring that individuals can access, modify, and delete their data. To achieve compliance, AI developers may need to implement technical and organizational measures, such as anonymizing or pseudonymizing personal data, implementing privacy by design and default, and conducting regular assessments to identify and mitigate potential privacy risks. Failure to comply with GDPR can result in severe financial and reputational consequences, including fines of up to €20 million or 4\% of the company's global annual revenue, whichever is higher \cite{EU-GDPR}.

This paper addresses a critical gap identified in existing system-level responsible-ai-by-design frameworks, as highlighted in \cite{Lu2022}. The gap pertains to the absence of specific patterns necessary to encompass all ethical principles in designing and developing AI systems.We propose a pattern collection approach to bridge this gap, introducing several new patterns that fill the identified gap.
Our pattern collection includes essential additions to cover the ethical principles required for designing AI systems. Notably, we introduce innovative design patterns such as Zero-Trust AI \cite{Tidjon2022}, AI Ethics Auto-Testing, AI Ethics Patching, and Ethical Drift Detection. These patterns address specific ethical considerations and contribute to the overall responsible design of AI systems.
To ensure adherence to ethical principles throughout the lifecycle of AI systems, we emphasize the significance of employing AI Ethics Orchestration and Automated Response (EOAR). This framework enables regular audits and testing of models, facilitating the comprehensive assessment of ethical compliance in AI systems.
By addressing this gap and incorporating these advancements, our paper aims to provide a comprehensive and practical framework for designing AI systems that are both ethical and lawful. Furthermore, it emphasizes integrating responsible design patterns (RDP) into the machine learning pipeline (ML pipeline) to create responsible AI systems.

The main contributions of this paper can be summarized as follows:
\begin{itemize}

\item we conducted a survey that focused on validating patterns such as Zero-Trust AI, AI Ethics Auto-Testing, AI Ethics Patching, Ethical Drift Detection, and AI Ethics Orchestration and Automated Response (EOAR) to gain insights from experts on the current implementation of responsible design patterns (RDP). 

\item We propose AI ethics patterns integrated with the ML pipeline as well as a Bias Mitigation design pattern to create responsible AI systems. Our framework offers a practical approach to ensuring AI systems' ethical and legal design.

\item We examine and validate responsible design patterns (RDP) for ML pipelines across industry expertise on data collection, which provides evidence of the effectiveness of the proposed framework in real-world scenarios.

\end{itemize}

The paper is structured as follows: Section~\ref{Background} provides an overview of the background, including AI ethics, responsible AI, ML design patterns, and responsible ML design patterns. Section~\ref{Related Work} presents recent related work on creating responsible AI systems.Section~\ref{Methodology} describes the methods used to conduct the survey and subsequent thematic analyses. Section~\ref{Results} presents the survey results. Section~\ref{Responsible AI design patterns} presents an overview of the proposed design pattern for Responsible AI. Section~\ref{Discussion and Limitations} provides a comprehensive discussion, suggestions for future work, and limitations of the study, followed by the conclusion in Section~\ref{Conclusion}.

\section{Background}\label{Background}

This section introduces the concept of AI ethics, discusses the importance of responsible AI, and highlights the role of machine learning (ML) design patterns in addressing ethical considerations in the ML pipeline. Specifically, ML design patterns provide solutions to the ethical challenges associated with developing, deploying, and using AI technologies, ensuring that AI is used responsibly and ethically. Furthermore, when ML design patterns are integrated into the ML pipeline, it enables the development and training of AI systems that tackle ethical issues such as bias and fairness, transparency and explainability, accountability and responsibility, and social and environmental benefits.

\subsection{AI Ethics}

AI ethics is a dynamic field that addresses the ethical implications of developing, deploying, and utilizing artificial intelligence (AI) technologies. The increasing presence of AI systems in our daily lives raises significant ethical concerns that must be tackled to ensure responsible and ethical AI practices \cite{jobin2019global, bostrom2018ethics,boddington2017towards }. One of the most crucial ethical issues in AI revolves around bias and fairness, as AI systems heavily rely on data, and biased data can result in biased outcomes, leading to unfair treatment of certain groups \cite{Ntoutsi2020,Shestakova2021}.

To handle machine learning (ML) processes and safeguard data privacy and security, public and private organizations generally adhere to a combination of legal and ethical policies. These policies are designed to comply with applicable data protection and privacy laws, such as the General Data Protection Regulation (GDPR) in Europe or the California Consumer Privacy Act (CCPA) in the United States, and minimize risks associated with ML system use (GDPR; CCPA). Organizations must also implement processes to ensure data privacy and address biases, including regular assessments to identify and mitigate potential biases in training data and ML system outputs. Ethical considerations are incorporated into the development and deployment of ML systems through internal guidelines, codes of conduct, and ethical impact assessments. Security policies, such as encryption, access controls, and data backup procedures, safeguard personal data from unauthorized access or use.

Organizations follow various practices to ensure data privacy, security, and ML performance. Regular audits, both internal and external, are conducted to review the security and privacy of data, internal systems, and software, including those related to ML. Monitoring and evaluating ML models' accuracy, reliability, and overall performance is crucial which involves regular testing, validation, and feedback loops to address any performance issues. Access to ML APIs is controlled through authentication and authorization mechanisms, such as OAuth or API keys, to protect the processed data and ensure authorized usage. Data quality is evaluated through assessments, validation processes, and continuous monitoring of data sources, pipelines, and storage systems. Moreover, organizations often establish Service Level Agreements (SLAs) for data and ML management, defining expectations, responsibilities, and requirements for performance, security, and privacy.

Overall, the policies, practices, and evaluations organizations adopt to handle ML processes and ensure data privacy and security depend on factors such as data nature, jurisdiction, and industry sector. However, compliance with relevant laws and regulations is essential, and implementing appropriate processes and procedures is crucial to minimize risks associated with ML system usage.

\subsection{Responsible AI}

Responsible AI  is the concept of developing and deploying AI systems that are ethical, transparent, fair, and trustworthy. It involves considering the potential impact of AI on individuals and society as a whole and taking steps to ensure that AI systems are developed and used responsibly \cite{Eitel-Porter2020,Lu2022a,Maalej2023}. AI systems should be transparent and explainable to ensure that users and stakeholders can make informed decisions about their use. In addition, AI systems should be designed and deployed reasonably that do not discriminate based on race, gender, or ethnicity. Privacy should also be a top priority, with personal information protected and not misused. Finally, those involved with AI development and use should be held accountable for its impact on individuals and society. 

Responsible AI practices can help ensure that AI technologies are developed and deployed ethically and responsibly \cite{Lu2022b}. So, ethical principles should be defined to guide the development of AI technologies. These principles may include fairness, accountability, transparency, and safety. Next, potential sources of bias in the data and algorithms used in AI systems should be identified and analyzed to ensure that they do not unfairly disadvantage any group of people. Continuously monitoring the performance of the AI system is also crucial to ensuring that it operates as intended and does not cause harm or negative impacts. Additionally, the decision-making process of the AI system should be transparent and explainable to build trust with users and stakeholders. Finally, privacy should also be a key consideration in the development of AI systems, with an emphasis on minimizing the collection and storage of personal data and ensuring that any data that is collected is protected and used only for its intended purpose \cite{Lu2022, Vakkuri2021, Khan2022, Daza2022}.
Evaluating the impact of the AI system on society, including any potential ethical implications, is critical to ensure that it meets the needs and expectations of users and stakeholders. Finally, ethical principles and practices should be regularly reviewed and updated to remain relevant and practical. By incorporating responsible AI practices into the AI development lifecycle, developers can help to ensure that AI technologies are used ethically and responsibly. This promotes the responsible use of AI technologies for the benefit of society and builds trust with users and stakeholders \cite{Mirbabaie2022}.

\subsection{ML Design Patterns}

\begin{figure*}
  \centering
  \includegraphics[width=\textwidth]{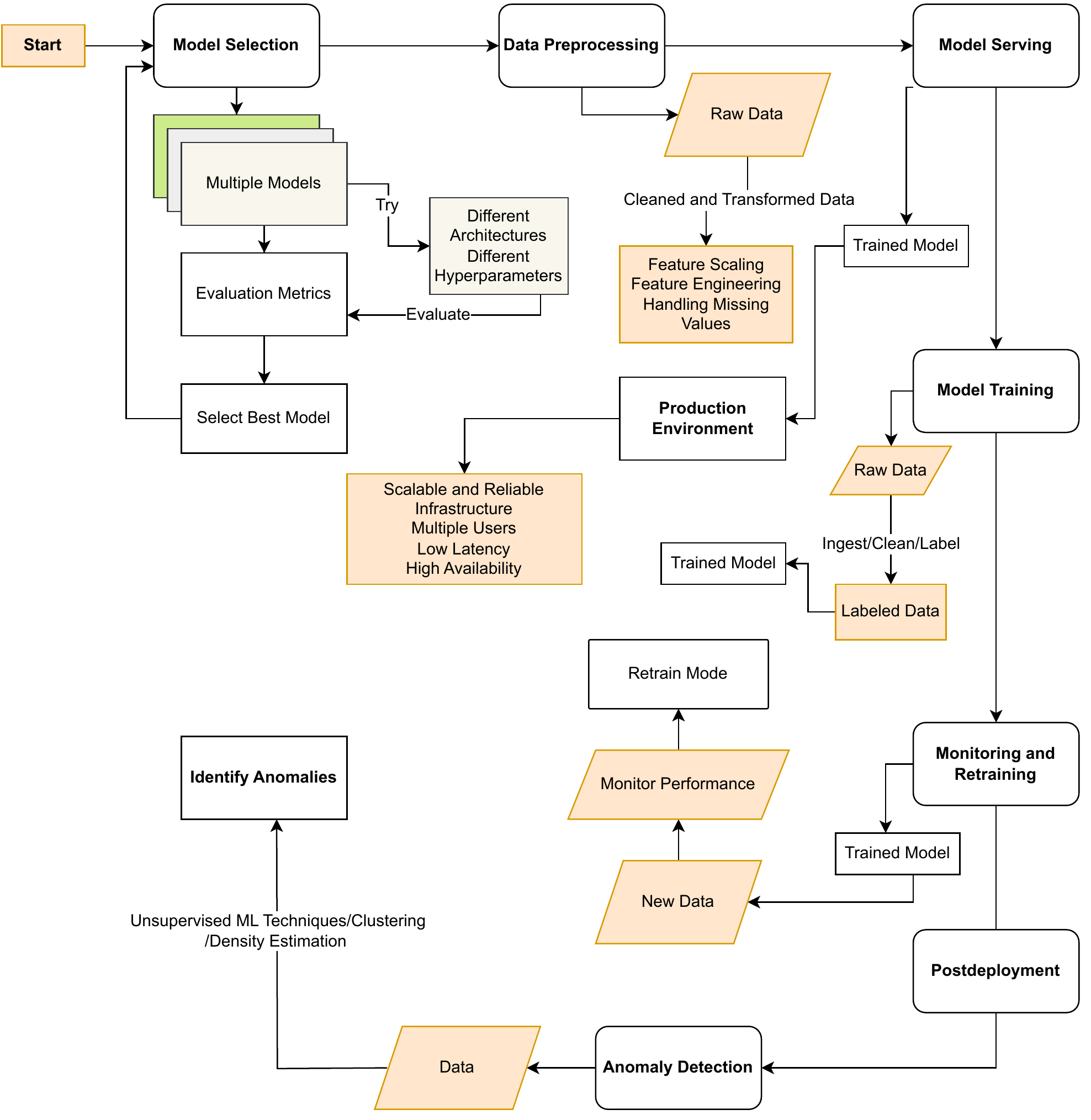}
  \caption{Machine Learning Design Patterns Life cycle}
  \label{fig:intro1}
\end{figure*}

Machine learning (ML) design patterns are reusable solutions to common problems in developing ML systems. In addition, design patterns help to ensure that the resulting system is efficient, maintainable, scalable, and robust \cite{ibm_blog,enisa_report}. Here are some commonly used ML design patterns as shown in Figure~\ref{fig:intro1}:

\textbf{\textit{Model Selection:}} Choosing the right model architecture and hyperparameters is essential for the success of a machine learning system. In order to attain peak performance, it is imperative to conduct thorough testing of different models employing varied architectures and hyperparameters. The model that exhibits superior evaluation metrics must be selected.

\textbf{\textit{Data Preprocessing:}} Data preprocessing involves cleaning, normalizing, and transforming data before it is fed into an ML model. In addition, the Data Preprocessing pattern involves applying techniques such as feature scaling, feature engineering, and handling missing values to ensure that the data is suitable for training an ML model.

\textbf{\textit{Model Serving:}} Model Serving involves deploying an ML model to a production environment where it can be used to make predictions. In addition, the Model Serving pattern involves setting up a scalable and reliable infrastructure that can handle requests from multiple users and ensure low latency and high availability.

\textbf{\textit{Model Training:}} Model Training involves using labelled data to train an ML model. The Model Training pattern involves a pipeline for ingesting, cleaning, and labelling data and an ML framework to train the model on the labelled data.

\textbf{\textit{Monitoring and Retraining:}} ML models can degrade over time due to changes in the data distribution or other factors. The Monitoring and Retraining pattern involves continuously monitoring the model's performance and retraining it on new data to maintain its accuracy and effectiveness.

\textbf{\textit{Anomaly Detection:}} Anomaly Detection involves identifying unusual or unexpected data points that do not conform to the regular pattern. The Anomaly Detection pattern involves using unsupervised ML techniques such as clustering and density estimation to identify anomalies in the data.

\subsection{Responsible ML Design Patterns}

Design patterns are commonly used in AI development to provide reusable solutions to common problems. Similarly, responsible design patterns can be used in machine learning pipelines to promote responsible AI and ensure that ethical considerations are considered. The concept of patterns in the design of responsible AI systems is gaining increasing attention in the AI research community. Patterns are commonly understood as reusable and proven solutions to common design problems. In the context of responsible AI, patterns are associated with the states or state transitions of a provisioned AI system, effectively guiding architects to design a responsible AI system \cite{ibm_blog,enisa_report}. One prominent example of such a pattern is the "Explainable AI" pattern, which addresses the need for AI systems to provide clear and understandable explanations for their decisions and actions. Another example is the "Fairness" pattern, which ensures that AI systems do not discriminate against specific individuals or groups based on protected characteristics such as race, gender, or age.

\section{Related Work}\label{Related Work}

The ethical use of artificial intelligence (AI) is a significant concern in AI ethics. It addresses ethical considerations related to developing, deploying, and using AI technologies. Responsible AI design patterns promote the ethical use of AI while limiting its irresponsible use. Furthermore, it recognizes the importance of developing AI and technology innovation responsibly and ethically, as highlighted by various studies \cite{jobin2019global,bostrom2018ethics,boddington2017towards}.

One of the most critical ethical issues in AI is bias and fairness. Biased AI can result in unfair treatment of particular groups of people. Therefore, it is essential to guarantee that AI systems are created and educated with impartial data to tackle this problem. Previous studies \cite{Tidjon,fjeld2020principled} have identified the need for trustworthy AI systems that meet specific trustworthy requirements, including fairness. These studies have reviewed and compared existing trustworthy AI approaches based on their trustworthiness.

Lu et al. \cite{Lu2022} conducted a recent study that thoroughly examines responsible AI design and offers practical advice on developing transparent, accountable, fair, and safe AI systems. The paper's incorporation of design patterns is especially valuable, as it provides developers with reusable solutions to common design challenges encountered in responsible AI design. Nevertheless, the proposed design patterns need critical components to monitor systems end-to-end. In related work, Lu et al. \cite{Lu2022c} emphasized the significance of responsible AI design and its role in building trust in AI systems. They introduced the concept of design patterns as reusable solutions to design problems in AI, providing a comprehensive set of design patterns for responsible AI design, including accountability, transparency, fairness, and safety. However, it is essential to note that some studies have highlighted the need to consider specific design patterns during the pre-processing stage, such as detecting and cleaning data, to mitigate vulnerabilities like data leakage. While the paper recognized the continuous nature of AI design and the need for ongoing improvement, it may not have adequately stressed the importance of early consideration of specific design patterns during the AI design process. 

Sanderson et al. \cite{Sanderson2023} addressed the gap between high-level AI ethics principles and practical techniques for designing and developing responsible AI systems. The paper conducted semi-structured interviews with researchers and engineers from Australia's national scientific research agency to examine how their practices align with a set of high-level AI ethics principles proposed by the Australian Government. However, while the paper provided valuable insights into the practical application of these principles, it did not sufficiently emphasize the need to integrate some ethically responsible AI compounds into the design of the ML pipeline. For instance, the paper did not consider the ethical implications of using biased training data, nor did it address the importance of regularly monitoring AI models for potential biases and unintended consequences. Furthermore, the paper did not cover some critical ethical considerations in the design of the ML pipeline, such as the importance of interpretability and the need to ensure that the system does not harm human rights or environmental well-being. Therefore, while the paper provided valuable insights into the practical implementation of high-level AI ethics principles, it needed to address some critical ethically responsible AI compounds that must be considered in designing the ML pipeline.

Peters, Dorian, et al. \cite{Peters2020} proposed the AI Ethics Principles and the Data Ethics Canvas as tools for ethical design practice when developing AI systems. While these frameworks help identify and address ethical issues related to data collection, processing, and usage, the paper has limitations. For example, it does not provide a comprehensive overview of all possible ethical considerations related to AI systems, and more complex ethical issues related to AI still need to be fully addressed, such as the potential impact on employment and societal structures. Despite these limitations, the paper's strengths lie in providing actionable frameworks for ethical design practice and emphasizing the importance of collaboration and multidisciplinary teams in the development process. However, ethical considerations should be integrated into all aspects of the ML pipeline, including pre-processing stages, model selection, and evaluation. This integration requires interdisciplinary collaboration between experts in AI, ethics, policy, and other related fields to ensure that all ethically responsible AI compounds are considered.

Recent work by Maalej, W., Pham, Y. D., and Chazette, L.\cite{Maalej2023} discusses the challenges of applying requirements engineering (RE) in machine learning (ML) projects based on several recent studies. The studies highlighted the difficulties encountered by practitioners in ML projects, including an insufficient understanding of customers or too many expectations. Additionally, practitioners found that requirements are more uncertain for ML systems than non-ML systems, and the goals often need to be more abstract. Unlike traditional systems, ML systems expose black-box behaviour that is hard to specify, analyze, and validate. One limitation of this work is that it mainly focuses on ML projects and only considers other AI projects. Additionally, the proposed aspects should be more extensively discussed and evaluated in the context of real-world projects, which may limit their practical applicability. 

Smith et al.\cite{Smith2020} proposed a framework for responsible AI design, focusing on ethical considerations in algorithm design and development. They emphasized the importance of avoiding biases, promoting fairness, and ensuring the ethical impact of AI systems. However, we extend their work by presenting a comprehensive framework covering all ML pipeline stages, including data pre-processing, model training, and model deployment. We also introduce new design patterns and emphasize using AI ethics orchestration and automated response (EOAR) for regular auditing and testing. Taking this approach ensures ethical principles are followed throughout the entire lifecycle of the AI system, leading to a more comprehensive and responsible design.

Based on our literature review, current research on responsible AI design provides valuable frameworks for ethical design practice when developing AI systems. While these frameworks may not cover all possible ethical considerations, they are valuable tools for identifying and addressing ethical issues during data and AI development.

In order to create AI systems that are ethical, reliable, and meet the needs of users and society, we suggest incorporating a set of new practices such as Zero-Trust AI, AI Ethics Auto-Testing, AI Ethics Patching, Ethical Drift Detection, and AI Ethics Orchestration and Automated Response (EOAR). These practices ensure that all ethical principles are covered when designing AI systems.

\section{Methodology}\label{Methodology}

The objective of conducting research on software engineering for responsible AI design patterns is to tackle the challenges that arise in the process of designing, developing, and deploying AI systems that are responsible, ethical, and trustworthy. The impact of AI systems on society can be significant, and their behaviour can influence individuals and groups in diverse ways, including perpetuating biases, discrimination, and privacy breaches.

This research aims to validate patterns such as Zero-Trust AI, AI Ethics Auto-Testing, AI Ethics Patching, Ethical Drift Detection, and AI Ethics Orchestration and Automated Response (EOAR) and to update the existing framework (responsible-ai-by-design) proposed by Lu et al. \cite{Lu2022}. Furthermore, it proposes a comprehensive framework incorporating responsible design patterns into Machine Learning (ML) pipelines to mitigate risks and ensure the ethical development of AI systems. This research strives to advance the development of responsible AI systems that align with human values and social norms, fostering trust and accountability in the AI system. Additionally, it aims to offer guidance and recommendations to stakeholders, including policymakers, software developers, and users of AI systems, on designing and deploying responsible AI systems.

The primary research question of this paper can be formulated as follows:

\textbf{What AI engineering practices and design patterns can be utilized to ensure responsible AI development?}

In the following, we will provide the steps taken to answer this research question along with the research findings and proposed bias mitigation design patterns.

\subsection{Search Strategy:}

\subsubsection{Survey}

To gain insights from experts on the current implementation of responsible design patterns (RDP), we conducted a survey that focused on validating patterns such as Zero-Trust AI, AI Ethics Auto-Testing, AI Ethics Patching, Ethical Drift Detection, and AI Ethics Orchestration and Automated Response (EOAR).As well,the survey explored current practices, processes, and tools related to AI ethics. 

We distributed the survey among organizations utilizing AI and experts in AI ethics and data management. The data collected from the survey were analyzed using a thematic analysis approach, which focused on experts' perceptions, current practices, and risk assessment tools related to AI ethics.

\subsubsection{Motivation and Intent}

The motivation behind the survey was to gain insights and perspectives from various stakeholders, including organizations that currently utilize AI and experts in AI ethics and data management. We aimed to understand the current practices, processes, and tools employed by experts in the field to identify risks in data management processes. We could identify patterns that require improvement and enhancement by gaining insights into their existing approaches. This information was crucial in identifying the specific patterns that needed to be integrated into ML pipelines to promote responsible and ethical AI development.

We sought to validate the proposed responsible design patterns through real-world scenarios. By examining and validating these patterns, we aimed to ensure their effectiveness and suitability in addressing the potential risks and challenges associated with AI systems.

This survey gathered valuable data and feedback from AI ethics and data management experts. This information guided us in identifying the specific patterns that should be integrated into ML pipelines, thereby contributing to developing a comprehensive framework for responsible AI system development.

The survey questions and proposed design patterns were developed based on the current literature and best practices in the field of AI ethics and data management as follows:

\begin{itemize}

\item{\textit{What are the current practices of managing data for AI ethics adoption in the public and private sectors?}}
\item{\textit{What current processes and tools are used for managing data and AI processes for adopting ethics in the public and private sectors?}}
\item{\textit{What tools are used by individuals and organizations involved in data and AI management processes to identify and mitigate risks?}}

\end{itemize}

The survey sought insights and perspectives from various stakeholders, including organizations that currently utilize AI and experts in AI ethics and data management. By conducting this survey and analyzing the data obtained using a thematic analysis approach, we aimed to identify the current tools for identifying risks in data management processes and, ultimately, to better understand the current state of managing data for AI ethics adoption in the public and private sectors.

\subsubsection{Target Experts}
The target experts for this survey would be individuals with expertise and knowledge in AI ethics and data management, including but not limited to academics, researchers, practitioners, and consultants who specialize in these areas. In addition, they may have experience working with organizations in the public and private sectors to develop and implement ethical frameworks for AI, as well as expertise in the legal, social, and technical aspects of AI ethics. These experts may be identified through professional networks, academic institutions, industry associations, and other relevant sources.

\subsubsection{Data Collection}
The survey link was shared on LinkedIn, and a call for participation was issued on our LinkedIn page, using relevant hashtags to reach individuals specializing in AI ethics. The survey collected data on various aspects, including organizational practices for AI ethics adoption, compliance standards, privacy and security enforcement in the ML pipeline, and risk assessment tools. In addition, open-ended questions were included to gather insights into internal procedures, compliance policies, ethics strategies, data privacy and bias management, security policies, audit processes, and service-level data and ML management agreements.

\subsection{Participants}

Based on the responses, the survey participants consisted of experts from various organizations and domains, highlighting the collaborative nature of AI ethics adoption and risk management. The organizations that participated in the survey included Ethically Aligned AI, Thomson Reuters, IBM, Conseil de l'innovation du Québec, Université de Sherbrooke, CSIRO (Commonwealth Scientific and Industrial Research Organization), Polytechnique Montréal, and the Ministry of Health - Quebec.

The participants' expertise spanned a wide range of domains related to AI and technology. These areas included AI ethics, AI adoption, data management, responsible AI, cybersecurity, law/tech regulation, software engineering, and technology transfer. The positions mentioned varied greatly, representing the diverse roles and responsibilities within the AI field. Some of the mentioned positions included AI security architect, CEO, legal and compliance for AI, data engineering advisor, innovation and AI adoption director, professor, principal research scientist, information security officer, and responsible AI (PhD).

The collaboration between experts from diverse organizations and domains is crucial in ensuring a comprehensive and multidisciplinary approach to AI governance and risk management. By involving experts from different fields, the survey highlights the importance of incorporating varied perspectives to address the ethical and responsible development and implementation of AI technologies. Furthermore, this interdisciplinary collaboration enables a holistic consideration of ethical principles, regulatory compliance, technical expertise, and societal implications, leading to more informed and well-rounded decision-making in AI development and deployment.

\section{Analysis}

\begin{figure*}
  \includegraphics[width=\textwidth]{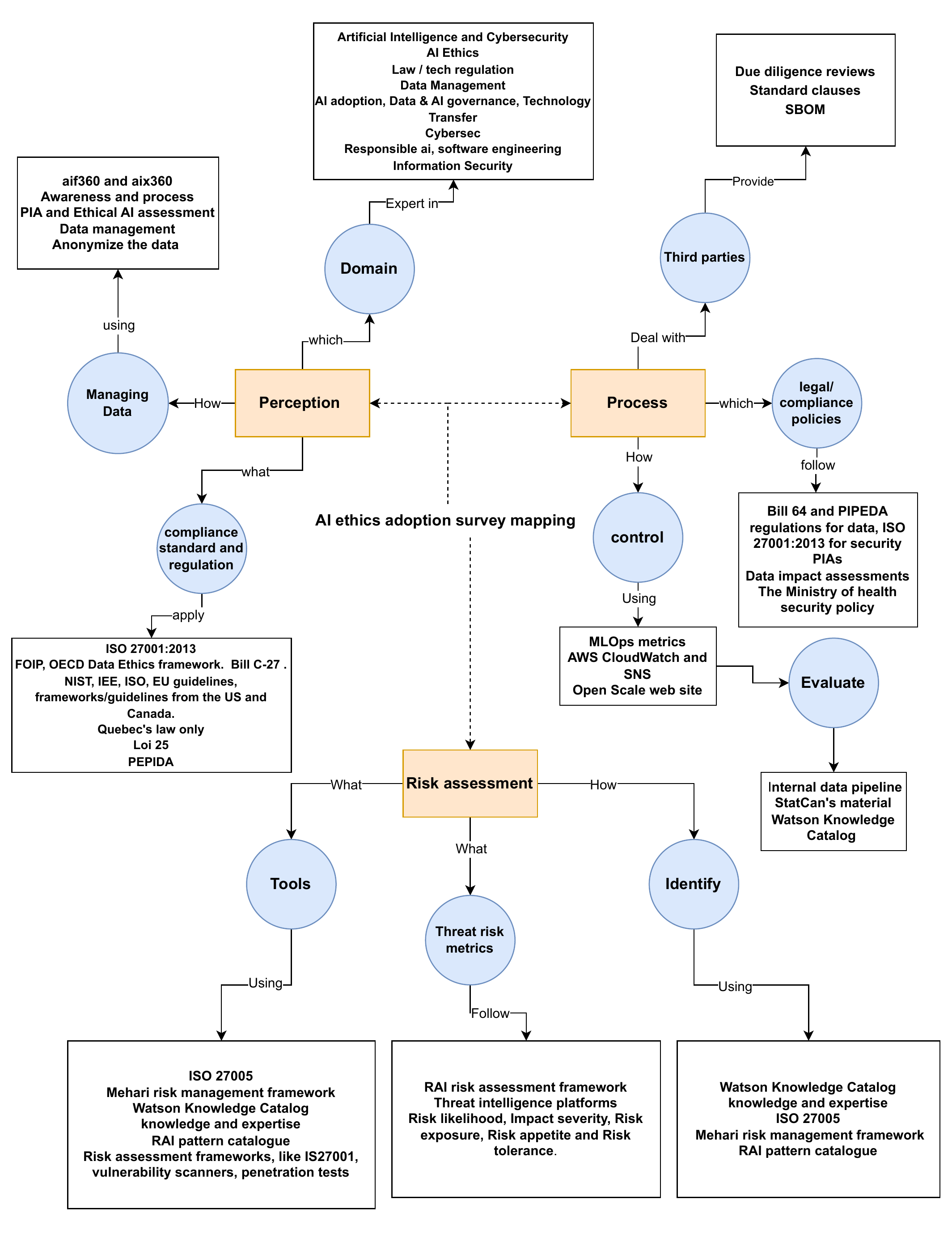}
  \caption{AI ethics adoption survey mapping}
  \label{fig:adopt3}
\end{figure*}

The survey analysis can be divided into three main parts, as shown in Figure~\ref{fig:adopt3}: experts' perceptions of artificial intelligence ethics, current practices in place, and risk assessment tools. In addition, the survey focused on significant factors and correlations related to how organizations handle personal and customer data about AI ethics adoption.

Participants were asked about their organization's practices for adopting AI ethics, compliance standards and regulations, and privacy and security enforcement in the machine learning (ML) pipeline. Additionally, the survey included open-ended questions about internal procedures for dealing with third parties, compliance policies for ML pipelines, data privacy and bias management, ethics strategies and policies, security policies for end-to-end data protection, audit processes for data, controlling ML access API performance, evaluating data quality, and service level agreements for data and ML management.

The third part of the survey focused on risk assessment, including current tools used to identify threat risks in the data management process and current threat risk metrics to control data and the ML pipeline. Overall, the survey results provide insights into how organizations approach AI ethics adoption and manage the associated risks.

\section{Results}\label{Results}

\subsection{Artificial Intelligence Ethics Perception}

This section aims to gather information on the perception of AI ethics, organizational readiness for AI ethics adoption, and awareness of security regulations and risks associated with implementing AI ethics practices.

\subsubsection{Expert Perception on Responsible Design Patterns for Machine Learning Pipelines}

The findings of the survey indicate that the participants possess a deep understanding of the intricacies and complexities involved in AI ethics and risk management, which can be leveraged to devise effective strategies and policies for the responsible development and deployment of AI technologies, specifically for responsible design patterns for Machine Learning pipelines.

There is general acknowledgement among the participants that AI ethics is a critical aspect of AI development and usage. Some participants are directly engaged with AI ethics, such as creating standards and policies to comply with ethical AI principles and researching global standards and best practices for ethically regulating AI. For instance, P2, the CEO of Ethically Aligned AI, mentioned, \textit{"I work with AI ethics directly. I develop standards, policies for compliance with ethical AI principles."} Similarly, P3 noted, \textit{"IBM sells a platform to help companies of all sizes in defining, monitoring and tuning ethical behaviour of their AI models."}.

Others emphasize that AI ethics is an essential consideration for the entire field of AI, and it is crucial for the social acceptance and adoption of AI systems. They also note that issues related to AI ethics have been a significant barrier to AI adoption. P4, the Conseil de l'innovation du Québec mentioned, \textit{"Issues related to AI ethics have been important obstacles to AI adoption."}

The proposed practice of developing and using AI to benefit humans while minimizing ethical risks aligns with the concept of responsible AI. Responsible AI aims to ensure that AI technologies are developed and used ethically, transparently, and accountable. This approach recognizes the potential of AI to bring significant benefits while also acknowledging the associated risks and the need to integrate ethical considerations into the development and use of AI systems.

\subsubsection{Current Practices}

This section aims to collect information about the current practices of adopting AI ethics. This practice includes compliance standards and regulations, privacy and security enforcement, and governance of the ML pipeline.

The survey found that many organizations have policies and procedures for AI ethics adoption, including compliance standards and regulations. However, implementing these policies varies across organizations, with some having more comprehensive policies and procedures than others. Many organizations also reported having policies and procedures in place for privacy and security enforcement in the ML pipeline, but some were still implementing these policies.

In terms of ML pipeline governance, many organizations reported having procedures for dealing with third parties, compliance policies for ML pipelines, and ethics strategies and policies. However, there were still gaps in some areas, such as data privacy and bias management, security policies for end-to-end data protection, audit processes for data, and controlling ML access API performance.

The survey revealed that organizations are progressing in adopting AI ethics and implementing policies and procedures to ensure responsible AI development and deployment. However, there is still room for improvement in some areas, and organizations should continue to develop and refine their policies and procedures to ensure the responsible use of AI technologies.

\begin{tcolorbox}[colback=lightgray]
The survey findings demonstrate a growing recognition among public and private sector organizations of the importance of ethical considerations in developing and using AI technologies, with a need for increased education and proactive adoption of ethical principles in data-handling practices.
\end{tcolorbox}

\subsection{Handle personal and customer data about AI ethics adoption}

The findings reveal that organizations exhibit varying degrees of awareness and adoption of ethical principles in their data handling. Some organizations reported utilizing open libraries such as AIF360 (AI Fairness 360)\footnote{\href{https://aif360.mybluemix.net/}{AIF360}} and AIX360 (AI Explainability 360)\footnote{\href{https://aix360.mybluemix.net/}{AIX360}} to address data bias and explain data. For instance, one of the participants mentioned that  \textit{"We use open libraries to remove bias on data and explain data"}. Others have separate privacy management and compliance programs for personal data and combine evaluations for privacy and ethical AI assessments.

However, there are also instances where regulatory compliance is the primary driver of AI ethics practices within organizations, indicating that there may be room for improvement in adopting ethical principles beyond regulatory mandates.

Notably, some respondents could not answer as they were not in charge of data management within their organization, highlighting the importance of cross-functional collaboration and communication in implementing AI ethics practices. For example, P4 mentioned, \textit{"I am not in charge of data management within my organization"}.

The study results suggest a need for increased awareness regarding AI ethics, particularly in areas such as fairness, bias, responsibility, and transparency. Organizations can benefit from adopting a more comprehensive approach to AI ethics that incorporates best practices in ethical principles and processes. Anonymizing data or utilizing other ethical methods can also serve as additional means to ensure the ethical handling of personal and customer data in AI adoption. P6 indicated,\textit{"Anonymize the data or other ways"}.

Overall, this study highlights the importance of prioritizing ethical considerations in developing and deploying AI systems and emphasizes the need for organizations to take a proactive approach to integrate ethical principles into their data-handling practices.

\subsection{Compliance Standards and Regulations}

The responses to the research survey regarding compliance standards and regulations followed by organizations in the context of AI ethics reveal a diverse range of approaches.

ISO 27001:2013 was mentioned as a commonly followed standard, focusing on information security management. Other standards and frameworks mentioned include NIST (National Institute of Standards and Technology)\footnote{\href{https://www.nist.gov/artificial-intelligence}{NIST}}, ISO (The International Organization for Standardization)\footnote{\href{https://www.iso.org/home.html}{ISO}}, EU (the European Commission) guidelines\footnote{\href{https://ec.europa.eu/commission/presscorner/detail/en/ip_22_6110}{EU(the European Commission)}}, and frameworks/guidelines from the US and Canada, suggesting a more comprehensive approach to AI ethics that encompasses various ethical considerations beyond information security.

Some respondents mentioned specific regulations like FOIP (The Freedom of Information and Protection of Privacy Act)\footnote{\href{https://www.alberta.ca/freedom-of-information-and-protection-of-privacy-overview.aspx}{Freedom of Information and Protection of Privacy (FOIP)}} in the public sector in Alberta, Quebec's Law 25, and proposed incoming regulation Bill C-27. These regulations focus on privacy and may require organizations to adhere to specific privacy requirements when handling personal data.

One respondent mentioned that their school may have adopted the best data privacy and security practices, such as data minimization techniques, regular privacy impact assessments, and adherence to international privacy laws and regulations like GDPR.

Depending on their industry and jurisdiction, organizations may follow different compliance standards and regulations. Adhering to specific regulations may be necessary, but organizations may benefit from adopting a more comprehensive approach to AI ethics encompassing various ethical considerations beyond regulatory compliance.

\subsection{Tools used to enforce privacy and security}

The survey results on the tools presently utilized to ensure privacy and security in the ML pipeline.
One respondent mentioned using the ARX tool for anonymization and privacy \footnote{\href{https://arx.deidentifier.org/}{ARX tool}}, ARX is a comprehensive open-source software for anonymizing sensitive personal data. While another mentioned using the Adversarial Robustness Toolbox (ART)\ tool for security \footnote{\href{https://github.com/Trusted-AI/adversarial-robustness-toolbox}{ART\ tool}}.\textit{" \textit{we use arx tool for anonymization and privacy. For security, we use ART}"} However, some respondents reported not using specific tools for privacy and security enforcement in the ML pipeline.

One respondent mentioned PIAs(Privacy Impact Assessments) as a tool for enforcing privacy and security in the ML pipeline, and another mentioned having a separate team conducting security reviews for all projects involving data and models.\textit{"\textit{We have a separate team that does security review for all projects using data and/or models}"}.

One respondent mentioned using IBM's Open Scale platform, which helps companies define, monitor, and tune the ethical behaviour of their AI models. This platform is often used with IBM Open Pages for model governance. Federated learning was also mentioned as a tool for privacy and security in the ML pipeline.

The responses indicate a need for further development and adoption of specific tools to enforce privacy and security in the ML pipeline. While some organizations may already be using specific tools or techniques, others may need to explore and implement more effective solutions to address privacy and security concerns in AI.

\subsection{Internal Procedures for Dealing with Third Parties}

Based on the responses to the survey question regarding internal procedures for dealing with third parties, 62.5 \% of the participants answered "Yes," while 37.5 \% answered "No."

For those who answered "Yes," three responses were provided regarding their internal procedures for dealing with third parties. The first response was the use of standard clauses. It is still being determined what these standard clauses refer to, but they are likely contractual terms and conditions commonly used in agreements with third parties.

Participants mentioned SBOM \footnote{\href{https://www.cisa.gov/}{SBOM}}, which stands for "Software Bill of Materials." An SBOM is a list of all software components used in a particular product or system, along with their versions, licenses, and known vulnerabilities. It identifies and manages potential security risks associated with third-party software components.

It also mentioned due diligence reviews. Due diligence reviews involve conducting a thorough investigation and analysis of a third party's background, reputation, financial stability, legal compliance, and other relevant factors to ensure they meet the necessary standards and requirements for working with the organization.

\subsection{Legal/compliance policies used to handle machine learning pipelines}

Several different responses were provided in response to the legal/compliance policies used to handle machine learning pipelines. The first response suggests the participant uses the Open Scale website for legal and compliance policies related to machine learning pipelines. It is unclear what policies or guidelines are provided on the Open Scale website, but it is a source of information for participants.

The survey's responses indicate that the participant uses no legal or compliance policies for machine learning pipelines. This response suggests that the participant may need to know the legal and compliance considerations relevant to machine learning pipelines or may have yet to implement policies specifically for this purpose.

Some response mentions specific regulations and standards that are followed for data privacy and security, including Bill 64 and PIPEDA\footnote{\href{https://www.priv.gc.ca/en/privacy-topics/privacy-laws-in-canada/the-personal-information-protection-and-electronic-documents-act-pipeda/}{PIPEDA}} for data and ISO 27001:2013 for security. These policies are designed to ensure compliance with legal and industry standards for data privacy and security in machine learning pipelines.

Other responses include "NA," "Not sure," "N/A," and "I don't have access," which suggest a lack of familiarity or access to relevant legal and compliance policies. One participant mentions that they are still working on implementing policies and are currently using privacy impact assessments (PIAs) and data impact assessments to evaluate the viability and compliance of their projects.
\begin{tcolorbox}[colback=lightgray]
The findings highlight the need for organizations to prioritize establishing and implementing comprehensive legal and compliance policies specifically tailored to machine learning pipelines to ensure adherence to relevant regulations and industry standards.
\end{tcolorbox}
\subsection{Legal/compliance for data privacy and bias management}

The survey responses regarding participants' approaches to legal/compliance for data privacy and bias management reveal a range of strategies and levels of implementation. For example, one participant stated compliance with ISO 27001:2013, Bill 64, and the Personal Information Protection and Electronic Documents Act (PIPEDA) for data privacy but did not have specific measures for addressing bias. Instead, they highlighted adherence to sector-specific data privacy regulations, such as the Freedom of Information and Protection of Privacy Act for the public sector and the Personal Information Protection Act (PIPA)\footnote{\href{https://www.alberta.ca/personal-information-protection-act.aspx}{PIPA}} in Alberta's private sector. Another response indicated the absence of definitive standards or tools for bias management, while a different participant mentioned utilizing bias management tools, conducting ethics evaluations, and incorporating mitigation measures. A third response emphasized treating data privacy separately and ensuring compliance based on the personal information involved. Finally, oneFinally, one participant mentioned relying on the Open Scale website for addressing legal/compliance aspects of data privacy and bias management, although the specific policies and guidelines provided on the website remained unclear. Additional responses included "N/A," "yes," and "not sure," indicating varying degrees of familiarity with and implementation of legal/compliance measures for data privacy and bias management.

\begin{tcolorbox} [colback=lightgray]
These findings underscore the need for organizations to carefully consider and implement appropriate legal and compliance measures for data privacy and bias management. It is essential to align with relevant regulations, develop tools and standards for bias management, and ensure a comprehensive approach that addresses data privacy and bias considerations in AI projects.
\end{tcolorbox}

\subsection{Security policies in place for end-to-end data protection}

As the use of data continues to increase in various sectors, it has become essential to have robust security policies in place to protect sensitive information. In this research survey, we sought to understand if organizations have security policies for end-to-end data protection and, if yes, how they are implemented.

Out of the total respondents, 75 \% answered positively, indicating that their organizations had security policies for end-to-end data protection. 12.5 \% answered in the negative, while 12.5 \% responded with a "\textit{maybe}" indicating some level of uncertainty.

Five respondents provided additional information on implementing their security policies for those who answered positively. For example, one participant mentioned that their city has a cybersecurity policy that specifies the details of the security measures to be taken. Another participant mentioned that their organization has comprehensive security policies that cover the data lifecycle at all stages.

Some respondents referred to IBM tools such as QRadar and Watson Knowledge Catalog, indicating the use of these tools to implement security policies. One respondent mentioned that their organization has a policy in place for data protection but did not provide further details.

For the 12.5 \% of respondents who responded negatively or with uncertainty, no additional information was provided to explain their responses.

\begin{tcolorbox}[colback=lightgray]
The research survey revealed varying approaches and levels of implementation in the public and private sectors regarding the tools and procedures used for managing data for AI ethics adoption, internal procedures for dealing with third parties, legal/compliance policies used to handle machine learning pipelines, legal/compliance for data privacy and bias management, and security policies in place for end-to-end data protection. While some organizations have implemented specific tools and policies, others need to explore and implement more effective solutions to address privacy and security concerns in the context of AI.
 \end{tcolorbox}

\subsection{Regular audit process in place}

The survey results indicate that 50\% of respondents have a regular audit process for data, internal systems, and software, while 25\% do not, and the remaining 25\% are still determining. Participants who responded positively have different approaches to auditing.

One participant stated that their city auditors run audits on these systems as part of their processes, while another believed the audit is conducted yearly. On the other hand, one respondent mentioned that their organization helps clients define and systematically apply an audit process through tools such as Open Pages \footnote{\href{https://www.ibm.com/products/openpages-with-watson?utm_content=SRCWW&p1=Search&p4=43700074359384728&p5=p&gclid=CjwKCAjwgqejBhBAEiwAuWHioCNpoHNF9xYmoaLqy9WzmC4JCdCLrbUiUcn2qN3oMc4IuSLhr7DUVhoCl_MQAvD_BwE&gclsrc=aw.ds}{IBM OpenPages}}, Open Scale \footnote{\href{https://www.ibm.com/watson?utm_content=SRCWW&p1=Search&p4=43700074359379226&p5=p&gclid=CjwKCAjwgqejBhBAEiwAuWHioFzdBX8fjYOMTIgBSYoH-K8D0FO7e0g18wnwy6YvJRJBrILHqf5MFhoCqKkQAvD_BwE&gclsrc=aw.ds}{Watson OpenScale}}, and IBM Data Privacy.

Regular auditing of data, internal systems, and software ensures that organizations meet security and compliance requirements. In addition, it helps identify potential risks and vulnerabilities in the system, which can be mitigated before they cause any harm. It is encouraging that half of the respondents have a regular audit process. However, for those who do not, it may be worth considering implementing one to strengthen their security posture.

\subsection{Controlling model performance in production}

An important finding from the research survey is the wide range of approaches to controlling model performance in production. Organizations employ various methods, such as monitoring execution time and accuracy using MLOps metrics and tools like AWS CloudWatch and SNS. In addition, some respondents emphasized the importance of continuous oversight from the design stage throughout the model deployment. Interestingly, there were differences in recommendations, with one respondent highlighting IBM Open Scale as a recommended tool, while another indicated a lack of specific controls or tools. Additionally, one respondent mentioned using multi-modal approaches, although further clarification is needed to understand the specific strategies  
\begin{tcolorbox}[colback=lightgray]
These findings suggest the need for organizations to adopt suitable control mechanisms and standardized practices for monitoring and managing model performance in production environments.
\end{tcolorbox}

\subsection{Evaluate the data quality}

An important finding from the research survey is that organizations employ diverse approaches and tools to assess the performance and quality of their machine-learning models and data. In addition, the evaluation methods vary across organizations, indicating a need for standardized practices in the field.

Some organizations rely on MLOps metrics, such as accuracy, execution time, and the average number of requests, to evaluate the performance of their models and API access. Others utilize established tools like IBM Open Scale and the Open Scale website. These tools provide insights into model performance and facilitate monitoring and evaluation.

Regarding data quality evaluation, organizations employ a range of techniques. For example, one respondent mentioned, "\textit{we have an internal data pipeline to collect, extract, preprocess, and store high-quality data following our data policies}" This highlights using an internal pipeline to ensure the quality of the collected data.

Another respondent stated, \textit{"There are quite a few good books on traditional data quality metrics. However, I like StatCan's material - https://www.statcan.gc.ca/en/data-quality-toolkit."} This indicates using external resources like StatCan's material to guide and inform data quality metrics.

Furthermore, organizations may have specific tools in place to manage data quality. For example, one respondent mentioned, "We have a data quality catalog and other tools to ensure data remains useful for its purposes." This statement demonstrates how a data quality catalog and other tools can be used to ensure that the data remains useful.

As one respondent mentioned, organizations may leverage resources such as the "Watson Knowledge Catalog website" to aid in data quality management.

The varied use of tools and approaches, including internal data pipelines, external resources, and dedicated data quality catalogs, highlights the significance of ensuring the performance and quality of machine learning models and data across different organizations. 
\begin{tcolorbox}[colback=lightgray]
The finding for this question suggests that organizations should adopt tailored evaluation methods based on their specific needs and resources. In addition, standardization and best practices in evaluating model performance and data quality could contribute to more reliable and robust machine learning systems.
\end{tcolorbox}

\subsection{Service Level Agreement (SLA)}

Research findings indicate that most surveyed organizations have implemented a Service Level Agreement (SLA) for managing data and machine learning. Specifically, 62.5\% of the respondents reported having an SLA in place, while 12.5\% responded negatively, and 25\% were unsure.

When asked to provide further details, only one respondent explained that they assist clients in defining and monitoring their SLAs. This observation suggests that some organizations may need a clearer understanding of the purpose and requirements of an SLA for data and ML management, or they may face challenges in implementing one due to factors such as limited expertise or resources.

Having a well-defined SLA is crucial for organizations to ensure that their data and ML management services meet agreed-upon performance, reliability, and availability standards. The absence of an SLA can lead to potential issues such as a lack of accountability, inadequate data quality, and non-compliance with regulatory requirements. 
\begin{tcolorbox}[colback=lightgray]
Highlighting this finding, it becomes evident that a lack of understanding or implementation of SLAs may give rise to various data and ML management challenges. Consequently, organizations should prioritize the establishment of clear and comprehensive SLAs to address issues related to accountability, data quality, and regulatory compliance, thus ensuring effective and efficient data handling practices.
\end{tcolorbox}

\subsection{Risk Assessment Tools}

This section aimed to gather information on the current tools used to identify risk in the data management process and risk metrics used to control data and the ML pipeline.

The survey found that many organizations use various tools to identify risks in the data management process, including data quality checks, data bias detection, and data lineage tracking. However, there is still a need for more comprehensive tools to manage risk in the ML pipeline, such as model explainability, model interpretability, and adversarial attack detection. The survey also found that many organizations use risk metrics to control data and the ML pipeline, such as false positives and negatives. However, there is still a need for more comprehensive metrics to measure fairness, transparency, and accountability in the ML pipeline.
\subsection{Tools and Methods for Identifying Threat Risks in Data Management Processes}
The present analysis is based on responses to investigate the tools and frameworks used to identify threat risks in the data management process. The survey results suggest various tools and frameworks organizations employ to identify and mitigate threats and risks. Some of the tools mentioned by respondents include ISO 27005 and Mehari risk management framework(Method for Harmonized Analysis of RISk) \footnote{\href{https://www.enisa.europa.eu/topics/risk-management/current-risk/risk-management-inventory/rm-ra-methods/m_mehari.html}{Mehari risk management}}.

Additionally, one respondent identified a data impact assessment as a useful tool, as it flagged risks and required expert review to ensure their mitigation.

Another respondent recommended using Watson Knowledge Catalog for data quality, Open Scale for ML behaviour and Ethics, and Data Privacy for privacy threats. \textit{P3, "We recommend Watson Knowledge Catalog for data quality, Open Scale for ML behaviour and Ethics, Data Privacy for threats to privacy"}. In contrast, one respondent cited knowledge and expertise as the primary tool for identifying threat risks. At the same time, another mentioned using an RAI(Responsible AI Pattern ) pattern catalogue \footnote{\href{https://research.csiro.au/ss/science/projects/responsible-ai-pattern-catalogue/}{RAI pattern catalogue}}.

\begin{tcolorbox}[colback=lightgray]
The findings of this question highlight the variety of methods organizations use to identify and mitigate threat risks in the data management process. It is evident that there is no one-size-fits-all approach, and different organizations rely on various tools and methods depending on their unique needs and circumstances. Therefore, further research is required to determine the effectiveness of these tools and frameworks and how they can be best used to improve the security of data management processes.
\end{tcolorbox}

\begin{figure*}
  \centering
  \includegraphics[width=\textwidth]{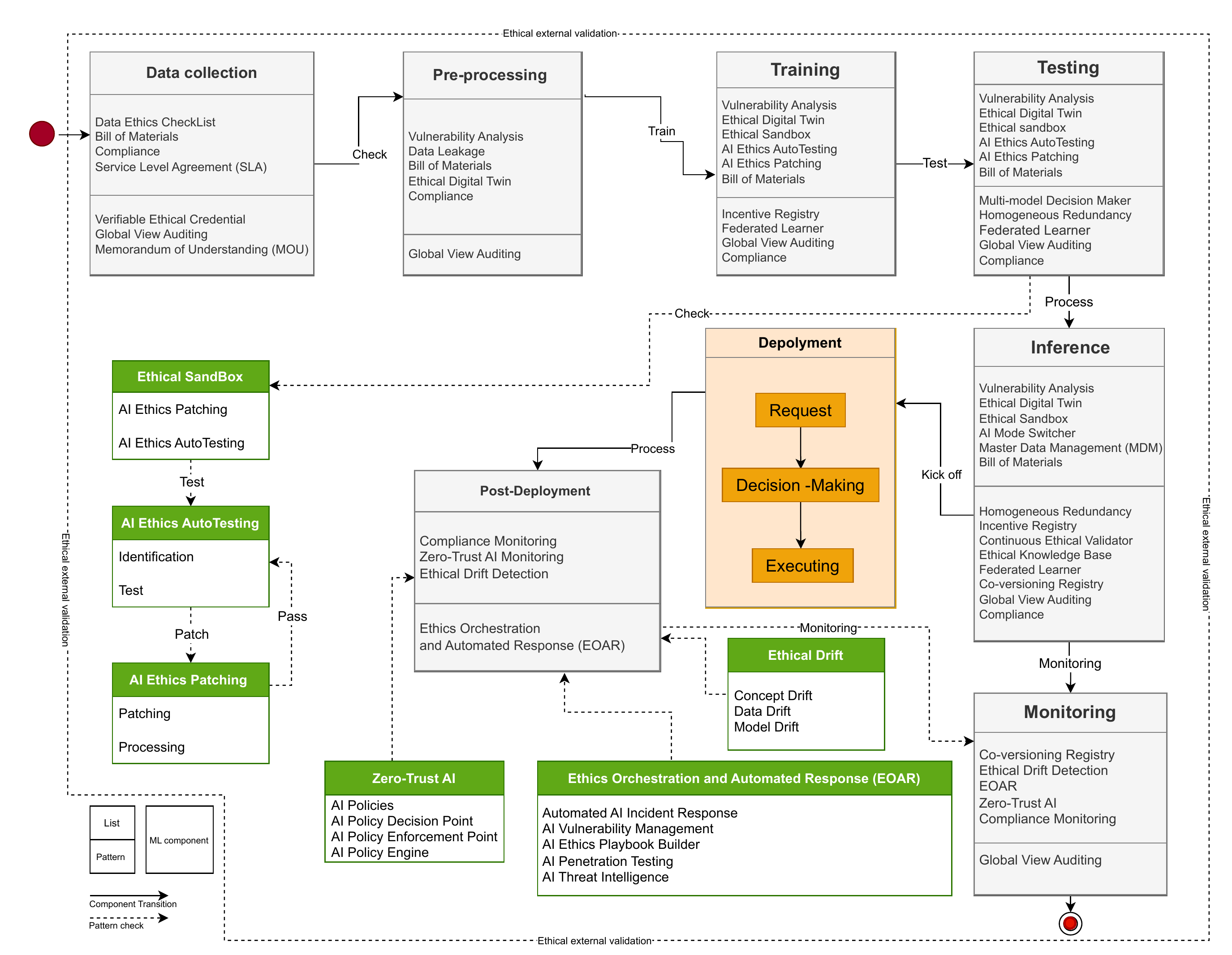}
  \caption{Responsible AI Design Patterns (RAI-DPs) Framework for Machine Learning Pipelines}
  \label{fig:patt1}
\end{figure*}

Organizations and individuals use various tools and frameworks to identify and reduce risk in data management processes. Respondents mentioned ISO 27005 and the Mehari risk management framework and data impact assessment as valuable tools. Some suggested Watson Knowledge Catalog for data quality, Open Scale for ML behaviour and ethics, and Data Privacy for privacy threats. However, some respondents emphasized knowledge and expertise as the primary tools for identifying threats and risks. The study suggests that there is no one-size-fits-all approach to managing data security risks, and additional research is needed to determine the effectiveness of these tools and frameworks.

\section{Responsible AI design patterns}\label{Responsible AI design patterns}

We developed a set of Responsible AI Design Patterns (RAI-DPs) that address the ethical concerns of AI systems, such as transparency, fairness, accountability, and explainability. These design patterns were developed based on the survey results, existing literature, and our team's software engineering and AI expertise. The proposed design patterns provide guidance and recommendations to ML engineers and scientists on designing and developing responsible AI systems that align with human values and social norms, thereby improving current practices.

Figure~\ref{fig:patt1} displays the Responsible AI Design Patterns (RAI-DPs) Framework for Machine Learning Pipelines. This diagram is crucial as it provides an overview of the overall structure and goals of the design patterns, allowing stakeholders to understand their relevance to their AI projects or initiatives. The framework includes the following components: Data collection, Pre-processing, Training, Testing, Inference, Deployment, Post-deployment, and Monitoring. In addition, it also features extension-related patterns for each ML component, such as Ethical sandbox, AI Ethics Auto-Testing, AI ethics patching, Zero-Trust AI, and AI Ethics Orchestration and Automated Response (EOAR). To clarify the concept behind this idea, we used a pattern template based on the extended pattern form \cite{meszaros1998pattern} and presented tables for some patterns. In the subsequent section, we will elaborate on each component in detail.

\subsection{Data Collection}

Data collection is a crucial stage in the machine learning (ML) pipeline, as it involves collecting and processing data used to train ML models. It is also essential for integrating ethical considerations into the AI system. Data quality is a critical aspect of data collection. Therefore, evaluating the data's quality is essential to ensuring its accuracy, completeness, and reliability. Furthermore, this evaluation helps identify any potential issues with the data and ensures its suitability for use in ML models\cite{danks2017data,mittelstadt2016ethics,buhrmester2011amazons,solaimani2021causal}.

To integrate responsible AI practices into the data collection stage, a Data Ethics Checklist shown in Table~\ref{tab:data-ethics-checklist}, Bill of Materials (BOM), and compliance shown in Table~\ref{tab:data-privacy-compliance} guidelines can be used to outline the hardware and software components used in the ML pipeline. A service level agreement (SLA) can also be used to define the level of service expected from the ML model\cite{ibm2021service}.

In the context of machine learning (ML) models, service level agreements (SLAs) are agreements between service providers and clients that specify the ML models' expected performance and quality metrics. SLAs can include various performance indicators, such as accuracy, speed, availability, and reliability, that the ML model must meet to satisfy the client's needs.

SLAs help ensure the ML models perform as expected and provide reliable client results. They can also facilitate ethical considerations in developing and deploying ML models by setting expectations for their performance and behaviour. For example, an SLA includes provisions that require the ML model to provide unbiased and fair decisions, protect user privacy, and comply with relevant laws and regulations (see Table~\ref{tab:service-level-agreement}).
In addition to these practices, a second level of ethical considerations can be implemented during the data collection stage, including Verifiable Ethical Credentials, Global View Auditing, and Memorandum of Understanding (MOU) Table~\ref{tab:memorandum-of-understanding}. Verifiable Ethical Credentials can be used to ensure that data is collected from reliable sources and is collected ethically. 
A Global View Auditing can be used to oversee the data collection process and ensure that it aligns with ethical considerations. Finally, a Memorandum of Understanding (MOU) can be used to establish the responsibilities and expectations of all parties involved in the data collection stage. 
\begin{table*}
\caption{Data Ethics Checklist}
\centering
\resizebox{\textwidth}{!}
{
\begin{tabular}{|p{2.5cm}|p{5cm}|p{4cm}|p{2.5cm}|p{3.5cm}|}
\hline
\textbf{Pattern Name} & \textbf{Context} & \textbf{Problem} & \textbf{Forces} & \textbf{Solution} \\
\hline
Data Ethics Checklist & Protecting privacy and ensuring unbiased data collection & Privacy rights, biased data, non-compliance & Privacy protection, bias avoidance, ethical compliance & Implement a data ethics checklist to ensure privacy protection, avoid bias, ensure ethical compliance, and enhance transparency and accountability in data collection \\
\hline
\textbf{Related Patterns} & \multicolumn{4}{p{14.5cm}|}{Data Ethics Governance Framework, Data Privacy Impact Assessment} \\
\hline
\textbf{Consequences} & \multicolumn{4}{p{14.5cm}|}{Enhanced Privacy and Confidentiality, Improved Data Quality, Ethical Compliance, Trust and Transparency} \\
\hline
\end{tabular}
}
\label{tab:data-ethics-checklist}
\end{table*}
\begin{table*}[htbp]
\caption{Data Privacy Compliance}
\centering
\resizebox{\textwidth}{!}
{
\begin{tabular}{|p{2.5cm}|p{5cm}|p{4cm}|p{2.5cm}|p{3.5cm}|}
\hline
\textbf{Pattern Name} & \textbf{Context} & \textbf{Problem} & \textbf{Forces} & \textbf{Solution} \\
\hline
Data Privacy Compliance & ML pipeline, sensitive data collection & Legal and ethical obligations, privacy breaches & Data privacy compliance measures& Implement data privacy compliance measures in the data collection phase of the ML pipeline \\
\hline
\textbf{Related Patterns} & \multicolumn{4}{p{14.5cm}|}{Data Anonymization, Informed Consent, Secure Data Storage} \\
\hline
\textbf{Consequences} & \multicolumn{4}{p{14.5cm}|}{Ensure legal compliance, mitigate risks, build trust} \\
\hline
\end{tabular}
}
\label{tab:data-privacy-compliance}
\end{table*}

\subsection{Pre-processing}
Pre-processing is an important stage in the Machine Learning (ML) pipeline, where the data is prepared and cleaned before it is used to train the ML models. This stage is also essential for integrating ethical considerations into the AI system. Two key aspects that can be integrated into the pre-processing stage are vulnerability analysis and data leakage prevention (see Table~\ref{tab:vulnerability-analysis-data-leakage}). Additionally, the Bill of Materials (BOM) can be used to ensure that ethical considerations are considered when selecting hardware and software components used in the ML pipeline.

\textit{Vulnerability Analysis:} Vulnerability analysis involves identifying potential vulnerabilities in the data and the ML models and taking steps to mitigate those vulnerabilities. For example, this analysis can include identifying biases in the data and taking steps to remove those biases, as well as testing the ML models for potential vulnerabilities such as adversarial attacks.

\textit{Data Leakage Prevention:} Data leakage prevention involves ensuring that sensitive data is not leaked or exposed during the pre-processing stage. This data can include steps such as anonymization and encryption of sensitive data and access controls to ensure that only authorized personnel can access the data.

\textit{Bill of Materials:} The Bill of Materials is a comprehensive list of all the hardware and software components used in the ML pipeline. By including ethical considerations in the BOM, organizations can ensure that the components used in the ML pipeline are secure, transparent, and compliant with relevant regulations. These regulations can include data privacy, transparency, and accountability. It can also help minimize the risk of vulnerabilities, data leakage, and other ethical concerns arising during the development and deployment of an AI system.

\subsection{Training}

In the training phase of the AI pipeline, ML models are trained on the pre-processed data to learn patterns and make predictions. Responsible AI training involves considering ethical sandbox, Global View Auditing, AI ethics Auto-Testing, AI ethics patching, and compliance patterns to ensure the trained model is fair, unbiased, and transparent.

The Ethical sandbox pattern is an essential method that AI developers must use to test their models before deploying them in the real world. It involves creating simulations of various scenarios to observe how the model behaves in each one, thus enabling them to detect any biases or ethical concerns. Furthermore, developers can identify and address potential issues to reduce their impact by testing the model in a controlled setting. Therefore, AI developers must prioritize this method to ensure their models are free from biases and ethical concerns.


The global view Auditing pattern in the training phase involves reviewing the ML models developed by various teams within the organization to identify potential biases or ethical issues that may not be apparent at the local level. This can involve analyzing the data used to train the model, the algorithms used to make predictions, and the outcomes of those predictions.

\begin{table*}
\caption{Service Level Agreement (SLA)}
\centering
\resizebox{\textwidth}{!}
{
\begin{tabular}{|p{2.5cm}|p{5cm}|p{4cm}|p{2.5cm}|p{3.5cm}|}
\hline
\textbf{Pattern Name} & \textbf{Context} & \textbf{Problem} & \textbf{Forces} & \textbf{Solution} \\
\hline
Service Level Agreement (SLA) & ML pipeline deployment, user expectations & Misunderstandings, legal issues & Performance assurance, resource allocation, customer satisfaction & Implement a Service Level Agreement (SLA) between the AI service provider and the users/clients \\
\hline
\textbf{Related Patterns} & \multicolumn{4}{p{14.5cm}|}{Model validation, Continuous monitoring, Data quality assurance } \\
\hline
\textbf{Consequences} & \multicolumn{4}{p{14.5cm}|}{Define performance requirements, enhance transparency }\\
\hline
\end{tabular}
}
\label{tab:service-level-agreement}
\end{table*}

\begin{table*}
\caption{Vulnerability Analysis \& Data Leakage}
\centering
\resizebox{\textwidth}{!}
{
\begin{tabular}{|p{2.5cm}|p{5cm}|p{4cm}|p{2.5cm}|p{3.5cm}|}
\hline
\textbf{Pattern Name} & \textbf{Context} & \textbf{Problem} & \textbf{Forces} & \textbf{Solution} \\
\hline
Vulnerability Analysis \& Data Leakage & data leakage risks & Privacy and security requirements, complexity & Vulnerability analysis, protection measures & Conduct a vulnerability analysis. Implement appropriate protection measures \\
\hline
\textbf{Related Patterns} & \multicolumn{4}{p{14.5cm}|}{Privacy-Preserving Techniques, Secure Data Transfer} \\
\hline
\textbf{Consequences} & \multicolumn{4}{p{14.5cm}|}{Identify vulnerabilities, enhance privacy and security} \\
\hline
\end{tabular}
}
\label{tab:vulnerability-analysis-data-leakage}
\end{table*}

\begin{table*}
\caption{AI Ethics Auto-Testing}
\centering
\resizebox{\textwidth}{!}
{
\begin{tabular}{|p{2.5cm}|p{5cm}|p{4cm}|p{2.5cm}|p{3.5cm}|}
\hline
\textbf{Pattern Name} & \textbf{Context} & \textbf{Problem} & \textbf{Forces} & \textbf{Solution} \\
\hline
AI Ethics Auto-Testing & Testing phase, automated ethical evaluation &  Manual testing limitations, bias, time constraints & Complexity, consistent evaluation, ethical compliance & Introduce an AI ethics auto-testing mechanism \\
\hline
\textbf{Related Patterns} & \multicolumn{4}{p{14.5cm}|}{Ethical Sandbox,AI Ethics Patching} \\
\hline
\textbf{Consequences} & \multicolumn{4}{p{14.5cm}|}{Effective ethical assessment, adherence to guidelines} \\
\hline
\end{tabular}
}
\label{tab:ai-ethics-Auto-Testing}
\end{table*}

\subsection{Testing}

The testing phase of AI development plays a vital role in promoting responsible AI practices. Several patterns can be utilized during this phase:

\textbf{\textit{AI Ethics Auto-Testing:}}
AI ethics auto-testing uses automated tools and techniques to evaluate the ethical implications of an AI system. This type of testing is becoming increasingly important to identify all AI-threatening scenarios as concerns about the ethical implications of AI continue to grow (see Table~\ref{tab:ai-ethics-Auto-Testing}).

The goal of AI ethics auto-testing is to identify any potential biases, unfairness, or other ethical issues that may be present in an AI system and to provide recommendations for addressing these issues. This type of testing can involve various techniques, including data analysis, algorithmic auditing, and scenario testing.

Some of the key benefits of AI ethics auto-testing include the following:

\begin{itemize}
    \item \textbf{Improved transparency}: By testing AI systems for ethical issues, organizations can increase transparency and accountability and build stakeholder trust.
    \item \textbf{Reduced risks:} By identifying and addressing potential ethical issues early on, organizations can reduce the risk of legal and reputational damage.
    \item \textbf{Improved decision-making:} By ensuring that AI systems are free from biases and other ethical issues, organizations can make better, more informed decisions.
\end{itemize}

Overall, AI ethics auto-testing is essential for ensuring that AI is developed and used responsibly and ethically.

\textbf{AI Ethics Patching:}
AI ethics patching refers to identifying and addressing ethical issues in an AI system after deployment. Just as software patches are used to fix bugs or security vulnerabilities in computer programs, AI ethics patching involves modifying an AI system to address ethical concerns.

The need for AI ethics patching arises because AI systems can be complex and difficult to understand or anticipate their ethical implications fully. Additionally, ethical concerns may only emerge once an AI system is deployed and used in real-world contexts. Therefore, it is essential to have a mechanism for addressing ethical issues that may arise after deployment.

Some common ethical issues that may require AI ethics patching include bias, privacy violations, and unintended consequences. For example, an AI system used to make hiring decisions may inadvertently discriminate against certain groups of applicants, which would require patching to address the bias. Similarly, an AI system that processes personal data may inadvertently violate privacy laws, requiring patching to address the issue.

AI ethics patching can involve various techniques, including data analysis, algorithmic auditing, and scenario testing. The goal is to identify and address ethical issues promptly and effectively to ensure that the AI system is being used responsibly and ethically.
  
\subsubsection{Incentive Registry}
This pattern promotes responsible behaviour by offering rewards or incentives to developers who adhere to ethical AI practices. It encourages the implementation of ethical considerations throughout the AI development process, rewarding developers for their responsible approach.

\subsubsection{Federated Learner}
The Federated Learner pattern facilitates collaborative learning by allowing AI models to be trained on decentralized data sources without compromising privacy. It ensures that sensitive data remains within the respective organizations while enabling the development of accurate and robust AI models.

\subsection{Inference}
The inference process is a critical aspect of AI's operation, but ethical concerns have arisen surrounding its decision-making, particularly in crucial areas such as healthcare, finance, and autonomous systems. To address these concerns, responsible inference patterns have been created to ensure that AI makes ethical decisions. This paper will thoroughly examine these patterns to ensure they are implemented correctly.

\textbf{\textit{Ethical Sandbox:}}

The ethical sandbox is a testing environment used for ethical inference testing. It enables the system to be tested for ethical decision-making to ensure it makes ethical decisions. In addition, the ethical sandbox can help identify and correct ethical issues before the system is released into the real world.

\textbf{\textit{AI Mode Switcher:}}

The AI mode switcher allows users to switch the AI system to different modes, such as ethical or performance, based on their preferences. This pattern allows for flexibility in the decision-making process and can improve stakeholder trust in the system.

\textbf{\textit{Master Data Management (MDM):}}

Master Data Management can ensure the accuracy and consistency of the data used in the inference process. MDM can help to ensure that the master data used by an AI system is accurate and consistent across all systems and processes. This pattern is important because any errors or inconsistencies in the data can lead to incorrect decisions by the AI system. For example, in the case of a predictive maintenance system that uses machine learning to predict when equipment will fail, MDM can help to ensure that the historical data used to train the machine learning model is accurate and consistent, regardless of where the data was collected or stored. This process can improve the accuracy of the predictions made by the AI system and reduce the risk of unexpected equipment failures.
\begin{table*}
\caption{Memorandum of Understanding (MOU)}
\centering
\resizebox{\textwidth}{!}
{
\begin{tabular}{|p{2.5cm}|p{5cm}|p{4cm}|p{2.5cm}|p{3.5cm}|}
\hline
\textbf{Pattern Name} & \textbf{Context} & \textbf{Problem} & \textbf{Forces} & \textbf{Solution} \\
\hline
Memorandum of Understanding (MOU) &  Data collocation phase, stakeholder alignment & Lack of clarity, conflicts, legal and ethical issues & Transparent practices, stakeholder alignment & Establish a Memorandum of Understanding (MOU) among the stakeholders involved in the data collocation phase \\
\hline
\textbf{Related Patterns} & \multicolumn{4}{p{14.5cm}|}{Service Level Agreement (SLA), Compliance Patterns} \\
\hline
\textbf{Consequences} & \multicolumn{4}{p{14.5cm}|}{Enhanced alignment, transparency, compliance} \\
\hline
\end{tabular}
}
\label{tab:memorandum-of-understanding}
\end{table*}

\textbf{\textit{Homogeneous Redundancy:}}

Homogeneous redundancy involves using multiple AI models to reduce the risk of bias and errors in decision-making. This pattern can help improve the accuracy and robustness of the system and reduce the risk of unintended consequences due to the system's biases or limitations.

\textbf{\textit{Incentive Registry:}}

The incentive registry pattern involves offering rewards for ethical decision-making. This pattern can motivate developers and users to make ethical decisions and help ensure that ethical considerations are given adequate attention in decision-making.

\textbf{\textit{Continuous Ethical Validator:}}

The continuous ethical validator is a real-time monitoring system that ensures AI systems make ethical decisions. This pattern can help to identify and correct ethical issues as they arise rather than after the fact.\\

\textbf{\textit{Ethical Knowledge Base:}}

The ethical knowledge base involves incorporating ethical principles and guidelines into the decision-making process of the AI system. This pattern can ensure that the system makes ethical decisions and improve stakeholder trust.

\textbf{\textit{Federated Learner:}}

The federated learner is a collaborative learning pattern that allows AI models to be trained on decentralized data without violating privacy. This pattern can help to improve the accuracy and robustness of the system while also protecting the privacy of individuals and organizations that provide data to the system.\\

\textbf{\textit{Co-Versioning Registry:}}

The co-versioning registry is a pattern that involves keeping track of changes made to the AI system and its datasets to ensure that it continues to operate within ethical boundaries. This process can help to identify and correct ethical issues as they arise and can improve stakeholder trust in the system.\\

\subsection{Deployment}

The Deployment phase of a Machine Learning (ML) pipeline is an essential step in deploying an AI system, and it is also a critical stage for ensuring that ethical considerations are taken into account throughout the entire workflow process. The ML pipeline typically includes three key components: request, decision-making, and execution. These components can be used to integrate ethical considerations into the Deployment phase of the ML pipeline as illustrated in Figure~\ref{fig:patt1}.

\subsubsection{\textbf{Request}} 

The request component of the ML pipeline involves receiving a request for a prediction or decision from the AI system. This request can come from various sources, including an end-user or another system. In addition, the request component can include ethical considerations such as data privacy and consent. For example, organizations can require that requesters provide explicit consent for their data to be used in the AI system and can ensure that the data is stored securely and used only for the purposes for which it was collected.

\subsubsection{\textbf{Decision-Making}} 

The decision-making component of the ML pipeline involves using the AI system to make a prediction or decision based on the input data provided in the request. This process is typically done using a trained model developed and tested during the development phase of the ML pipeline. In addition, the decision-making component can include ethical considerations such as fairness and transparency. This process can involve ensuring that the AI system does not unfairly discriminate against certain groups of people and that the decision-making process is transparent and explainable.

\subsubsection{\textbf{Execution}} 

The ML pipeline's executing component involves providing the AI system's output back to the requester. This process can include a prediction, a decision, or other relevant information. To integrate ethical considerations into the ML pipeline, these components can be modified to include ethical decision-making criteria. In addition, the executing component can be modified to include ethical considerations such as accountability and safety. Organizations can ensure that the AI system is accountable for its actions and safe for end-users and other systems to use. Finally, the production phase of the ML pipeline can involve ethical considerations to ensure that AI systems are built and used responsibly and ethically.

\begin{table*}
\caption{Zero-Trust AI}
\centering
\resizebox{\textwidth}{!}
{
\begin{tabular}{|p{2.5cm}|p{5cm}|p{4cm}|p{2.5cm}|p{3.5cm}|}
\hline
\textbf{Pattern Name} & \textbf{Context} & \textbf{Problem} & \textbf{Forces} & \textbf{Solution} \\
\hline
Zero-Trust AI & Post-deployment, continuous ethical monitoring & Dynamic AI systems, ethical drift detection & Continuous monitoring, Automated response mechanism & Implement Zero-Trust AI, which involves the following tasks: AI Policies, AI Policy Decision Point, AI Policy Enforcement Point, AI Policy Engine \\
\hline
\textbf{Related Patterns} & \multicolumn{4}{p{14.5cm}|}{Zero-Trust AI Monitoring, Ethics Orchestration and Automated Response (EOAR)} \\
\hline
\textbf{Consequences} & \multicolumn{4}{p{14.5cm}|}{Enhanced monitoring, timely response, improved compliance} \\
\hline
\end{tabular}
}
\label{tab:zero-trust-ai}
\end{table*}

\subsection{Post-Deployment}
The post-deployment phase is crucial for ensuring the ongoing responsible and ethical behavior of AI systems after the deployment phase. It involves compliance monitoring, addressing ethical drift detection, and implementing  ethics orchestration and automated response mechanisms (EOAR). The following components are integral to this phase:

\subsubsection{Zero-Trust AI Monitoring}
Zero-Trust AI Monitoring involves the continuous monitoring of AI systems to ensure compliance with ethical policies following the proposed Zero-Trust AI model in \cite{Tidjon2022}. This includes the implementation of:

\begin{itemize}
    \item \textbf{AI Policies:} Defining and enforcing policies that govern the ethical behavior of the AI system.
    \item \textbf{AI Policy Decision Point:} Managing the decision-making process to ensure adherence to ethical policies.
    \item \textbf{AI Policy Enforcement Point:} Enforcing policies and ensuring that the AI system operates within ethical boundaries.
    \item \textbf{AI Policy Engine:} Overseeing the policy management process and facilitating ethical decision-making.
\end{itemize}

\subsubsection{Ethical Drift Detection}

Ethical drift detection in machine learning (ML) refers to a model or system trained on a specific set of ethical principles or values that slowly moves away from those principles or values over time as it is retrained on new data or deployed in new contexts. This issue can lead to unintended consequences, such as bias, discrimination against certain groups, or other ethical violations.

One of the leading causes of ethical drift in ML is the use of biased or unrepresentative data to train models. For example, if the training data is not diverse enough or is biased towards certain groups or outcomes, the resulting model may also be biased or discriminatory. In addition, using algorithms to make decisions in high-stakes contexts, such as hiring, lending, or criminal justice, can have significant consequences if those algorithms are not designed with ethical considerations.

In order to prevent ethical issues in machine learning (ML), it is crucial to fully grasp the ethical principles and values that should be followed during the development and deployment of ML systems. These principles should be integrated into every stage of the machine learning process, including data collection, ethics checklist, quality evaluation, and model development and deployment. There are three main aspects involved in this process.

\begin{itemize}
    \item \textbf{Concept Drift:} Monitoring and detecting changes in the underlying concepts and assumptions of the AI system's training data that may impact its ethical behaviour.
    \item \textbf{Data Drift:} Continuously evaluate the incoming data to identify any shifts that may introduce biases or ethical concerns.
    \item \textbf{Model Drift:} Monitoring the deployed AI models' performance and behaviour to detect deviations from the desired ethical standards.
\end{itemize}

\subsubsection{AI Ethics Orchestration and Automated Response (EOAR)}

The AI Ethics Orchestration and Automated Response (EOAR) can be used to audit regularly and test models to detect and address ethical issues~\cite{Tidjon2022}.

The automated response can take immediate corrective action when ethical drift is detected. In contrast, Ethics Orchestration can be used to ensure that ethical considerations are integrated into every stage of the ML development process. This process may include ongoing education and training for developers and stakeholders to ensure that ethical considerations remain prioritized throughout the entire ML development process.
Ethics Orchestration and Automated Response involve the automated management and response to ethical incidents and vulnerabilities. This process  includes the implementation of the following:

\begin{itemize}
    \item \textbf{Automated AI Incident Response:} Developing protocols and automated responses to address ethical incidents and ensure timely resolution.
    \item \textbf{AI Vulnerability Management:} Identifying and addressing vulnerabilities in the AI system that may lead to ethical concerns.
    \item \textbf{AI Ethics Playbook Builder:} Creating visual playbooks that outline steps to handle various ethical scenarios and incidents.
    \item \textbf{AI Penetration Testing:} Conduct penetration testing to identify potential vulnerabilities and assess the AI system's resilience against ethical attacks.
    \item \textbf{AI Threat Intelligence:} Gathering intelligence on potential threats and ethical risks in the AI system's operational environment.
\end{itemize}
 
\subsubsection{Zero-Trust AI}

The Zero-Trust model for AI systems can also promote ethical responsibility in addition to security. With the policies and rules managed by the Policy Decision Point (PDP), Policy Enforcement Point (PEP), and Policy Engine, organizations can ensure that their AI systems are developed and used in a responsible and ethical manner~\cite{Tidjon2022}. For example, the policies and rules managed by the PDP, PEP, and Policy Engine can include ethical considerations such as fairness, transparency, accountability, and respect for privacy. These policies and rules can be used to ensure that the AI system does not unfairly discriminate against certain groups of people, that it is transparent in its decision-making processes, that it is accountable for its actions, and that it respects the privacy of individuals whose data is being processed.

In addition, the Zero-Trust pattern shown in Table~\ref{tab:zero-trust-ai} can be used to ensure that ethical considerations are considered throughout the entire lifecycle of the AI system. The proposed Zero-Trust AI (ZTA) model, described by Tidjon et al. \cite{Tidjon2022}, incorporates EthicsOps to ensure the application and control of trustworthy AI principles from development and testing to deployment and maintenance. By using a combination of access controls, policy management, and authentication and authorization mechanisms, organizations can ensure that their AI systems are secure and compliant with relevant regulations while being developed and used responsibly and ethically.

The Zero-Trust model can be an effective way to promote both security and ethical responsibility in AI systems. It ensures that access to resources and services is controlled based on a set of policies and rules that are designed to promote both security and ethical considerations.

\subsection{Monitoring}

The monitoring phase of AI development is the last critical step to ensure that the system operates ethically with includes various patterns that can be applied to monitor the system's operation and decision-making process. The Co-Versioning Registry is one such pattern. It involves keeping track of changes made to the AI system and its datasets to ensure that it continues to operate within ethical boundaries. In addition, this pattern can help to prevent any unintended changes that could result in unethical decision-making.

The Ethical Drift Detection pattern involves monitoring the AI system for changes that could lead to unethical decision-making. This pattern emphasizes the need for continuous system monitoring to detect any changes that could result in ethical violations. It is essential to catch any changes early to prevent harm. 

Compliance is another critical pattern ensuring the ethical operation of AI systems. Compliance involves adhering to relevant regulations and ethical principles to ensure the AI system operates within ethical boundaries.

The Global-View Auditing pattern emphasizes the importance of providing stakeholders with a global view of the AI system's operation and decision-making process. This pattern can ensure transparency and accountability by providing stakeholders with an auditing view of the AI system's operation, thereby building trust and increasing confidence in the system.

\begin{figure*}
  \centering
  \includegraphics[width=\textwidth]{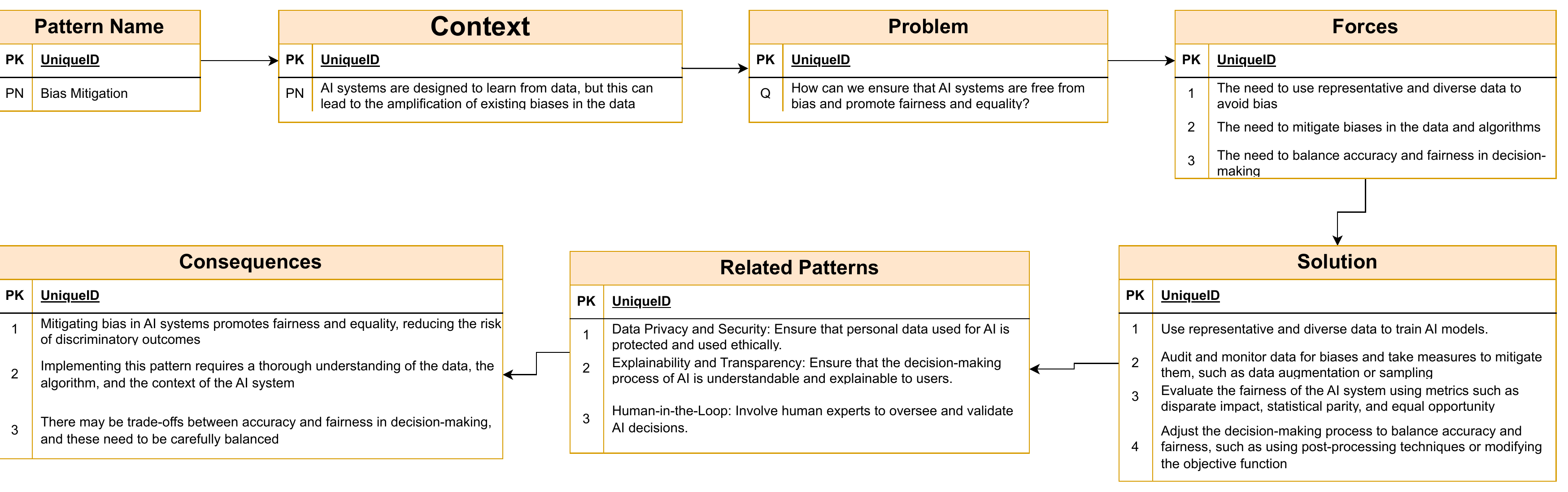}
  \caption{\footnotesize Bias Mitigation Example }
  \label{fig:bias1}
\end{figure*}

\section{Bias Mitigation Design Pattern for Responsible AI Systems }

In Figure~\ref{fig:bias1}, we present the Bias Mitigation pattern as an example of a responsible AI design pattern. The pattern aims to ensure that AI systems are free from bias and promote fairness and equality. We will discuss the context, problem, forces, solution, related patterns, and consequences of the Bias Mitigation pattern as follow the template of design patterns for responsible AI in~\cite{Lu2022}. Bias is a significant problem in AI systems, as they learn from data, and biases present in the data can amplify and result in unfair treatment and discrimination. The Bias Mitigation pattern aims to ensure that AI systems are free from bias and promote fairness and equality. The pattern has three leading solutions. Firstly, using representative and diverse data to train AI models ensures that the model's outputs are not biased toward certain groups. Secondly, auditing and monitoring data for biases and taking measures to mitigate them, such as data augmentation or sampling, helps reduce the risk of biased outputs. Finally, evaluating the fairness of the AI system using metrics such as disparate impact, statistical parity, and equal opportunity ensures that the system is not unfair toward any particular group.
The pattern's implementation requires a thorough understanding of the AI system's data, algorithm, and context. There may be trade-offs between accuracy and fairness in decision-making, which must be carefully balanced. To achieve this balance, post-processing techniques or modifying the objective function may be necessary.

The Bias Mitigation pattern is related to other patterns, such as Data Privacy and Security, which ensures that personal data used for AI is protected and used ethically. In addition, explainability and Transparency ensure that the decision-making process of AI is understandable and explainable to users. Finally, human-in-the-Loop involves human experts overseeing and validating AI decisions.

\section{Discussion and Limitations} \label{Discussion and Limitations}

\subsection{Discussion} 
Integrating Responsible AI Design Patterns (RAI-DPs) into the ML pipeline brings several benefits to developing ethical and legal AI systems. Patterns such as the Data Ethics Checklist and Data Privacy Compliance address critical issues related to privacy, fairness, and transparency in data collection and storage. By implementing these patterns, organizations can protect individuals' privacy, mitigate bias in data collection, and comply with ethical guidelines and regulations.

The proposed patterns enforced in the design include Service Level Agreement (SLA) and Memorandum of Understanding (MOU) patterns to help establish clear expectations and responsibilities between stakeholders involved in AI development and deployment. In addition, these patterns ensure that ethical considerations are explicitly addressed and agreed upon, promoting transparency and accountability in AI systems.

Vulnerability Analysis and AI Ethics Auto-Testing patterns introduce mechanisms for identifying and mitigating vulnerabilities and ethical risks in AI models. These patterns automate the process of testing and evaluating AI models for ethical considerations, reducing manual effort, time consumption, and potential human bias or oversight. Furthermore, this automation enables consistent and comprehensive ethical testing of AI systems, enhancing their overall ethical behaviour.

The Zero-Trust AI pattern emphasizes the importance of a holistic approach to security in AI systems. While technology tools like QRadar and Watson Knowledge Catalog play a crucial role in implementing security policies, they should be complemented with comprehensive security policies that cover the entire AI pipeline. This holistic approach ensures end-to-end protection and addresses the limitations of relying solely on technology for security.

Future work in responsible AI design patterns (RAI-DPs) can explore several avenues. Refining and expanding the set of RAI-DPs can provide a more comprehensive framework for addressing ethical issues in AI development. Considering the ethical implications of emerging AI technologies and incorporating them into the pattern set can help tackle novel ethical challenges.

Empirical studies are needed to assess the effectiveness of the proposed framework in different domains and contexts. Evaluating real-world AI systems developed using the RAIP-DPs against established ethical principles can provide valuable insights into the practical impact of the framework.

Integrating the RAI-DPs with other frameworks and methodologies for ethical AI development, such as the IEEE Global Initiative's framework, can lead to a more comprehensive and synergistic approach. Furthermore, this integration can leverage the strengths of different frameworks and provide organizations with a robust toolkit for designing and developing ethical AI systems.

In summary, integrating responsible AI design patterns (RAI-DPs) into the ML pipeline contributes to developing ethical and legal AI systems. These patterns address critical ethical considerations and challenges, automate ethical testing, and promote transparency, accountability, and end-to-end protection. Future research can further enhance the pattern set, evaluate its effectiveness in real-world scenarios, and explore integration with other ethical AI frameworks.

\subsection{Limitations}

This study has several limitations that should be acknowledged. Firstly, the relatively low number of responses to the survey may have limited the scope and depth of insights gained. Furthermore, although efforts were made to reach a diverse range of participants, the sample size might only partially represent the broader population involved in AI ethics and data management. This limitation could impact the findings' generalization and the identified patterns' comprehensiveness.

A second phase of the study is planned to address this limitation and expand upon the findings. This phase will involve in-depth interviews with AI ethics and data management experts. Engaging with a smaller group of participants in a more detailed manner will allow for a deeper exploration of current practices, processes, and tools used in identifying risks in data management processes. It will also provide additional insights into areas for improvement and potential new patterns that may have yet to be captured in the initial survey.

Another limitation to consider is the evolving nature of AI technology and ethical considerations. While the proposed framework covers a comprehensive range of ethical principles, new ethical challenges may emerge as AI technology advances. It is essential to recognize that the framework may need to be updated and adapted over time to address these emerging ethical challenges effectively. Continued monitoring and iteration of the framework will ensure its relevance and effectiveness in promoting ethical and legal AI systems.

While this study provides valuable insights into integrating responsible AI design patterns into the ML pipeline, its limitations should be acknowledged. The planned second phase of the study and ongoing monitoring and adoption of the framework will help address these limitations and contribute to the continuous improvement of ethical AI development practices.

\section{Conclusion}\label{Conclusion}

In this work, we proposed new AI ethics patterns for ML pipelines and a bias mitigation pattern to ensure trustworthy AI systems' development, deployment, and post-deployment. Experts in real-world scenarios have validated the proposed patterns to ensure their effectiveness. We introduced new design patterns, such as Zero-Trust AI, AI Ethics Auto-Testing, AI Ethics Patching, and Ethical Drift Detection, to cover all ethical principles for designing legal, ethical, and robust AI systems. We highlighted the importance of using AI Ethics Orchestration and Automated Response (EOAR) to audit and test models to address all ethical principles regularly. Furthermore, we presented a comprehensive and practical framework for designing ethical and legal AI systems by integrating AI ethics patterns into the ML pipeline. Our framework provides a practical and comprehensive approach to ensure the development of responsible AI systems. We conducted a survey to confirm these patterns and gained valuable insights into current practices for implementing Responsible Design Patterns (RDP). This survey improved our research and made our proposed framework more effective.


\section*{Acknowledgments}

We want to express our sincere gratitude to Dr. Qinghua Lu for providing valuable feedback on our work. Her insights and suggestions have greatly contributed to the improvement of our paper.

\bibliographystyle{ieeetr}
\bibliography{paper}

\begin{thebibliography}{10}

\bibitem{Alshehhi2022}
K.~Alshehhi, A.~Cheaitou, and H.~Rashid, ``Adoption frameworks for artificial
  intelligence in the public sector: A systematic review of literature,''

\bibitem{Daza2022}
M.~T. Daza and U.~J. Ilozumba, ``{A survey of AI ethics in business literature:
  Maps and trends between 2000 and 2021},'' {\em Frontiers in Psychology},
  vol.~13, 2022.

\bibitem{DeSanctis2020}
M.~{De Sanctis}, A.~Bucchiarone, and A.~Marconi, ``{Dynamic adaptation of
  service-based applications: a design for adaptation approach},'' {\em Journal
  of Internet Services and Applications}, vol.~11, no.~1, 2020.

\bibitem{Lu2022d}
Q.~Lu, L.~Zhu, X.~Xu, J.~Whittle, and Z.~Xing, ``{Towards a Roadmap on Software
  Engineering for Responsible AI},'' {\em Proceedings - 1st International
  Conference on AI Engineering - Software Engineering for AI, CAIN 2022},
  no.~i, pp.~101--112, 2022.

\bibitem{nabavi2022five}
E.~Nabavi and C.~Browne, ``Five ps: Leverage zones towards responsible ai,''
  {\em arXiv preprint arXiv:2205.01070}, 2022.

\bibitem{Ryan2021}
M.~Ryan and B.~C. Stahl, ``{Artificial intelligence ethics guidelines for
  developers and users: clarifying their content and normative implications},''
  {\em Journal of Information, Communication and Ethics in Society}, vol.~19,
  pp.~61--86, mar 2021.

\bibitem{EU2021}
E.~Commission, ``Proposal for a regulation of the european parliament and of
  the council laying down harmonized rules on artificial intelligence
  (artificial intelligence act),'' 2021.

\bibitem{IEEE2018}
R.~Chatila, K.~Firth-Butterfield, and J.~C. Havens, ``Ethically aligned design:
  A vision for prioritizing human well-being with autonomous and intelligent
  systems version 2,'' tech. rep., UNIVERSITY OF SOUTHERN CALIFORNIA LOS
  ANGELES, 2018.

\bibitem{Lu2022b}
Q.~Lu, L.~Zhu, X.~Xu, J.~Whittle, D.~Zowghi, and A.~Jacquet, ``Responsible ai
  pattern catalogue: a multivocal literature review,'' {\em arXiv preprint
  arXiv:2209.04963}, 2022.

\bibitem{Vakkuri2021}
V.~Vakkuri, K.~K. Kemell, M.~Jantunen, E.~Halme, and P.~Abrahamsson, ``{ECCOLA
  — A method for implementing ethically aligned AI systems},'' {\em Journal
  of Systems and Software}, vol.~182, p.~111067, 2021.

\bibitem{Khan2022}
A.~A. Khan, S.~Badshah, P.~Liang, M.~Waseem, B.~Khan, A.~Ahmad, M.~Fahmideh,
  M.~Niazi, and M.~A. Akbar, ``{Ethics of AI: A Systematic Literature Review of
  Principles and Challenges},'' {\em ACM International Conference Proceeding
  Series}, pp.~383--392, 2022.

\bibitem{saudi-ai-ethics}
S.~P. Agency, ``Saudi data and artificial intelligence authority announces ai
  ethics principles for public consultation.'' \url{https://www.spa.gov.sa},
  September 14 2022.

\bibitem{EU-GDPR}
{European Commission}, ``{Regulation (EU) 2016/679 of the European Parliament
  and of the Council of 27 April 2016 on the protection of natural persons with
  regard to the processing of personal data and on the free movement of such
  data, and repealing Directive 95/46/EC (General Data Protection
  Regulation)}.''
  \url{https://eur-lex.europa.eu/legal-content/EN/TXT/?uri=CELEX%3A32016R0679},
  2018.

\bibitem{Lu2022}
Q.~Lu, L.~Zhu, X.~Xu, and J.~Whittle, ``{Responsible-AI-by-Design: A Pattern
  Collection for Designing Responsible AI Systems},'' {\em IEEE Software},
  pp.~1--7, 2023.

\bibitem{Tidjon2022}
L.~N. Tidjon and F.~Khomh, ``Never trust, always verify: a roadmap for
  trustworthy ai?,'' {\em arXiv preprint arXiv:2206.11981}, 2022.

\bibitem{jobin2019global}
A.~Jobin, M.~Ienca, and E.~Vayena, ``The global landscape of ai ethics
  guidelines,'' {\em Nature Machine Intelligence}, vol.~1, no.~9, pp.~389--399,
  2019.

\bibitem{bostrom2018ethics}
N.~Bostrom and E.~Yudkowsky, ``The ethics of artificial intelligence,'' in {\em
  Artificial intelligence safety and security}, pp.~57--69, Chapman and
  Hall/CRC, 2018.

\bibitem{boddington2017towards}
P.~Boddington, {\em Towards a code of ethics for artificial intelligence}.
\newblock Springer, 2017.

\bibitem{Ntoutsi2020}
E.~Ntoutsi, P.~Fafalios, U.~Gadiraju, V.~Iosifidis, W.~Nejdl, M.~E. Vidal,
  S.~Ruggieri, F.~Turini, S.~Papadopoulos, E.~Krasanakis, I.~Kompatsiaris,
  K.~Kinder-Kurlanda, C.~Wagner, F.~Karimi, M.~Fernandez, H.~Alani, B.~Berendt,
  T.~Kruegel, C.~Heinze, K.~Broelemann, G.~Kasneci, T.~Tiropanis, and S.~Staab,
  ``{Bias in data-driven artificial intelligence systems—An introductory
  survey},'' {\em Wiley Interdisciplinary Reviews: Data Mining and Knowledge
  Discovery}, vol.~10, no.~3, pp.~1--14, 2020.

\bibitem{Shestakova2021}
V.~Shestakova, ``{Best Practices to Mitigate Bias and Discrimination in
  Artificial Intelligence},'' {\em Performance Improvement}, vol.~60, no.~6,
  2021.

\bibitem{Eitel-Porter2020}
R.~Eitel-Porter, ``{Beyond the promise: implementing ethical AI},'' {\em AI and
  Ethics}, vol.~1, pp.~73--80, oct 2021.

\bibitem{Lu2022a}
Q.~Lu, L.~Zhu, X.~Xu, and J.~Whittle, ``Responsible-ai-by-design: A pattern
  collection for designing responsible ai systems,'' {\em IEEE Software}, 2023.

\bibitem{Maalej2023}
W.~Maalej, Y.~D. Pham, and L.~Chazette, ``Tailoring requirements engineering
  for responsible ai,'' {\em Computer}, vol.~56, no.~4, pp.~18--27, 2023.

\bibitem{Mirbabaie2022}
M.~Mirbabaie, A.~B. Brendel, and L.~Hofeditz, ``{Ethics and AI in Information
  Systems Research},'' {\em Communications of the Association for Information
  Systems}, vol.~50, no.~1, pp.~726--753, 2022.

\bibitem{ibm_blog}
I.~Research, ``A comprehensive guide to responsible ai design patterns.''
  \url{https://www.ibm.com/blogs/research/2021/07/ai-design-patterns/}, 2021.
\newblock Accessed on March 27, 2023.

\bibitem{enisa_report}
{European Union Agency for Cybersecurity}, ``Pattern-based approaches to
  responsible ai.''
  \url{https://www.enisa.europa.eu/topics/ai-ml-and-big-data/responsible-ai/pattern-based-approaches-to-responsible-ai},
  2021.
\newblock Accessed on March 27, 2023.

\bibitem{Tidjon}
L.~N. Tidjon and F.~Khomh, ``The different faces of ai ethics across the world:
  A principle-to-practice gap analysis,'' {\em IEEE Transactions on Artificial
  Intelligence}, 2022.

\bibitem{fjeld2020principled}
J.~Fjeld, N.~Achten, H.~Hilligoss, A.~Nagy, and M.~Srikumar, ``Principled
  artificial intelligence: Mapping consensus in ethical and rights-based
  approaches to principles for ai,'' {\em Berkman Klein Center Research
  Publication}, no.~2020-1, 2020.

\bibitem{Lu2022c}
Q.~Lu, L.~Zhu, X.~Xu, J.~Whittle, D.~Douglas, and C.~Sanderson, ``{Software
  engineering for Responsible AI: An empirical study and operationalised
  patterns},'' {\em Proceedings - International Conference on Software
  Engineering}, no.~1, pp.~241--242, 2022.

\bibitem{Sanderson2023}
C.~Sanderson, D.~Douglas, Q.~Lu, E.~Schleiger, J.~Whittle, J.~Lacey,
  G.~Newnham, S.~Hajkowicz, C.~Robinson, and D.~Hansen, ``{AI Ethics Principles
  in Practice: Perspectives of Designers and Developers},'' {\em IEEE
  Transactions on Technology and Society}, pp.~1--1, 2023.

\bibitem{Peters2020}
D.~Peters, K.~Vold, D.~Robinson, and R.~A. Calvo, ``{Responsible AI—Two
  Frameworks for Ethical Design Practice},'' {\em IEEE Transactions on
  Technology and Society}, vol.~1, no.~1, 2020.

\bibitem{Smith2020}
J.~Smith, L.~Johnson, and M.~Davis, ``A framework for responsible ai design,''
  {\em IEEE Transactions on Artificial Intelligence}, vol.~8, no.~4,
  pp.~678--692, 2020.

\bibitem{meszaros1998pattern}
G.~Meszaros and J.~Doble, ``A pattern language for pattern writing,'' {\em
  Pattern languages of program design}, vol.~3, pp.~529--574, 1998.

\bibitem{danks2017data}
D.~Danks, A.~J. London, and C.~Weil, ``Data quality for data science,
  predictive analytics, and big data in supply chain management: An
  introduction to the problem and suggestions for research and applications,''
  {\em Big Data and Cognitive Computing}, vol.~1, no.~1, pp.~1--20, 2017.

\bibitem{mittelstadt2016ethics}
B.~D. Mittelstadt, P.~Allo, M.~Taddeo, S.~Wachter, and L.~Floridi, ``The ethics
  of algorithms: Mapping the debate,'' {\em Big Data \& Society}, vol.~3,
  no.~2, p.~2053951716679679, 2016.

\bibitem{buhrmester2011amazons}
M.~Buhrmester, T.~Kwang, and S.~D. Gosling, ``Amazon mechanical turk: A new
  source of inexpensive, yet high-quality, data?,'' {\em Perspectives on
  psychological science}, vol.~6, no.~1, pp.~3--5, 2011.

\bibitem{solaimani2021causal}
S.~Solaimani, M.~Zhang, and J.~Pathak, ``Causal inference in machine learning:
  An overview of methods and their implications for data quality and bias,''
  {\em Decision Support Systems}, vol.~145, p.~113489, 2021.

\bibitem{ibm2021service}
IBM, ``Service level agreements: What they are and how to use them.''
  \url{https://www.ibm.com/cloud/learn/service-level-agreements-slas}, 2021.

\end{thebibliography}

\restoregeometry

\end{document}